\pdfoutput=1
\documentclass[a4paper,11pt]{article}
\usepackage{jcappub} % for details on the use of the package, please
\usepackage[T1]{fontenc} % if needed
\usepackage[utf8]{inputenc}
\usepackage{nicefrac}
\usepackage{amsfonts}
\usepackage{lmodern}
\usepackage{bbold}
\usepackage{amsthm,graphicx}
\usepackage{epsfig}
\usepackage[normalem]{ulem}
\usepackage{latexsym,amssymb} 
\usepackage{amsmath,mathtools}
\usepackage{xcolor,xspace}
\usepackage{soul}
\usepackage{siunitx}
\usepackage{booktabs} % For professional looking tables
\usepackage{empheq}
\usepackage{cleveref}
\usepackage{enumerate}
\usepackage{enumitem}
\usepackage{algorithm}
\usepackage{algpseudocode}
\usepackage{arydshln}
\usepackage{float}
\usepackage{makecell}
\newfloat{algorithm}{t}{lop}
\newcommand{\zc}{\ensuremath{\mathcal{Z}}}

\newcommand{\red}[1]{{#1}}

\newcommand{\D}{\mathrm{d}}

\newcommand{\rev}[1]{{#1}}

\usepackage{color}

\notoc

\title{Bayesian optimisation for Bayesian evidence (\texttt{BOBE}) – a fast and efficient likelihood emulator for model selection}

\author[a]{Nathan Cohen}
\author[a]{Jan Hamann}
\author[b]{Ameek Malhotra}
\affiliation[a]{Sydney Consortium for Particle Physics and Cosmology, School of Physics, The University of New South Wales, Sydney NSW 2052, Australia}
\affiliation[b]{Physics Department, Swansea University, Swansea SA28PP, United Kingdom}

\emailAdd{nathan.cohen@unsw.edu.au}
\emailAdd{jan.hamann@unsw.edu.au}
\emailAdd{ameek.malhotra@swansea.ac.uk}

\abstract{The formalism of Bayesian model selection provides a very elegant way of ranking different physical models in terms of how compatible they are with a given set of observed data.  However, its practical application is often hampered by the challenge of having to compute the Bayesian evidence – a multi-dimensional integral over the product of likelihood and prior probability.  This usually necessitates a large number of function calls to the likelihood, which may become prohibitive in case of ``slow'', costly to evaluate likelihoods.  A possible solution to this problem lies in approximating the slow full likelihood by a fast emulated likelihood.
In this paper, we introduce \texttt{BOBE} (Bayesian Optimisation for Bayesian Evidence), a method to construct a Gaussian Process Regression (GPR)-based emulator.  \texttt{BOBE} utilises a Bayesian Optimisation algorithm designed specifically to (i) provide a realistic estimate of the emulator's uncertainty and its impact on the evidence calculation, and (ii) minimise the number of likelihood evaluations required in order to meet a given evidence accuracy goal.  We apply it to a number of toy examples as well as actual cosmological likelihoods, and demonstrate that training the emulator to a sufficient accuracy takes a factor of $\mathcal{O}(10^3)$ fewer direct likelihood evaluations than would be needed if one were to directly compute the evidence integral via nested sampling. \texttt{BOBE}'s overhead is independent of the likelihood computation time $t_\mathcal{L}$, making it particularly useful for ``expensive'' likelihoods with $t_\mathcal{L} \gtrsim 1$~s.  \texttt{BOBE} is written in \texttt{Python}, supports MPI parallelisation, takes advantage of automatic differentiation and just-in-time-compilation provided by \texttt{JAX}, can straightforwardly be implemented with cosmological data analysis frameworks such as \texttt{Cobaya}, and is available for download from \url{https://github.com/Ameek94/BOBE}.}

\begin{document}
\begin{flushright}
\large \tt{CPPC-2026-02}
\end{flushright}

\maketitle
\flushbottom

\newpage

\section{Introduction}
At the heart of the empirical sciences lies mankind's desire to understand Nature by describing reality in terms of quantitative theories, i.e., mathematical models.  Knowledge can be advanced via the conception of new models on the theory side, or by comparing, ranking or possibly rejecting models in light of new observations.  In either case, it is imperative to have rigorous, quantifiable criteria for model selection such that competing theories can be judged fairly.  

The framework of Bayesian statistics provides such a criterion, with the fundamental ingredient for Bayesian model comparison being the Bayesian evidence $\mathcal{Z}$~\cite{Trotta:2005ar,Trotta2008,Trotta:2017wnx}. For any given physical model $\mathcal{M}$ with free parameters $\theta$ and data $\mathcal{D}$, it is defined as the integral over the product of the likelihood $\mathcal{L}(\mathcal{D}|\theta,\mathcal{M})$ of the data and the prior probability $\pi(\theta|\mathcal{M})$ of the parameters:
\begin{align}
    \label{eq:evidence}
    \begin{split}
     \mathcal{Z} \equiv P(\mathcal{D}|\mathcal{M}) &=  \int \D \theta\; \mathcal{L}(\mathcal{D}|\theta,\mathcal{M}) \,\pi(\theta|\mathcal{M})\\
     &\equiv \int \D \theta\; \hat{\mathcal{P}}(\theta|\mathcal{M}).
     \end{split}
\end{align}
Note that the evidence $\mathcal{Z}$ is also the normalisation factor of the posterior probability of the parameters, $\mathcal{P}(\theta|\mathcal{M}) \equiv \tfrac{1}{\mathcal{Z}} \, \mathcal{L}(\mathcal{D}|\theta,\mathcal{M}) \,\pi(\theta|\mathcal{M})$, so one can interpret the integrand of Eq.~\eqref{eq:evidence} as an unnormalised posterior $\hat{\mathcal{P}}(\theta|\mathcal{M})$. 
If we assume two models $\mathcal{M}_1$ and $\mathcal{M}_2$ with equal model prior probabilities $P(\mathcal{M}_1) = P(\mathcal{M}_2)$, the ratio of two different models' evidences, also known as the Bayes factor,
\begin{align}
    \mathcal{B}_{12} \equiv \frac{\mathcal{Z}_1}{\mathcal{Z}_2} = \frac{ P(\mathcal{D}|\mathcal{M}_1)}{P(\mathcal{D}|\mathcal{M}_2)}\, ,
\end{align}
has a natural interpretation in terms of ``betting odds'' between the models.  Unlike frequentist approaches based on maximum likelihood values which only measure the quality of fit at a point, the evidence quantifies how well the model as a whole explains the data.  It rewards risky predictions that agree well with the data and punishes generic models which need parameters to be fine-tuned to achieve a good fit, thereby organically implementing the spirit of Occam's razor.
Bayes factors are commonly mapped to a strength of the evidence in favour of one model via Jeffreys' scale~\cite{jeffreys_scale,Trotta2008} (see, e.g., Tab.~\ref{tab:jeffreys_scale}).  
\begin{table}[h!]
    \centering
    \begin{tabular}{|c|c|c|} \hline
    $\ln \mathcal{B}_{12}$ & $\mathcal{B}_{12}$ & Evidence in favour of $\mathcal{M}_1$ \\ \hline 
    $\ln \mathcal{B}_{12} < 0$  & $\mathcal{B}_{12} < 1$ & none ($\mathcal{M}_2$ preferred) \\
    $0 \leq \ln \mathcal{B}_{12} < 1$  & $1 \leq \mathcal{B}_{12} \lesssim 3$ & inconclusive \\
    $1 \leq \ln \mathcal{B}_{12} < 2.5$  & $3 \lesssim \mathcal{B}_{12} \lesssim 12$ & weak \\ 
    $2.5 \leq \ln \mathcal{B}_{12} < 5$  & $12 \lesssim \mathcal{B}_{12} \lesssim 150$ & moderate \\ 
    $\ln \mathcal{B}_{12} \geq 5$  & $\mathcal{B}_{12} \gtrsim 150$ & strong \\ \hline
    \end{tabular}
    \caption{Jeffreys' scale, adapted from Ref.~\cite{Trotta2008}.}
    \label{tab:jeffreys_scale}
\end{table}

If one aims to reliably differentiate between different categories on the Jeffreys scale, it follows that the computation of the logarithm of the Bayes factors has an explicit precision requirement of $\mathcal{O}(0.1)$  (which similarly applies to the precision of the log-evidences, $\sigma_{\ln \mathcal{Z}}$).

Generally, the bottleneck of Bayesian model selection lies in the computation of the evidence integral (Eq.~\eqref{eq:evidence}).\footnote{
Evaluating the integrand of Eq.~\eqref{eq:evidence} involves three separate computations: (i) priors $\pi(\theta|\mathcal{M})$ (this is typically trivial), (ii) theory (i.e., the theoretical prediction of observables given parameters $\theta$ and model $\mathcal{M})$ and (iii) likelihood function (which takes as input the observables computed in step (ii)). }  Even for basic typical model selection problems encountered in cosmology, efficient state-of-the-art integration methods based on nested sampling~\cite{Skilling2006} (as implemented in packages such as \texttt{MultiNest}~\cite{Feroz2009}, \texttt{Polychord}~\cite{Handley2015a,Handley2015b}, \texttt{Dynesty}~\cite{Speagle2020} or \texttt{nautilus}~\cite{nautilus}), or using Markov-chain Monte Carlo (MCMC)-generated posterior samples~\cite{neal1998annealedimportancesampling,Heavens:2017afc,Kass01061995,mcewen2023machinelearningassistedbayesian,Polanska:2024arc,Lin:2025xbw}, reaching an evidence precision of $\sigma_{\ln \mathcal{Z}} \sim 0.1$ necessitates upwards of $10^5$-$10^6$ evaluations of the integrand~\cite{Sorensen:2025ywu} – a substantially larger number than would be required for parameter inference~\cite{LewisBridle2002, Foreman_Mackey_2013, brinckmann2018montepython3boostedmcmc}.

The number of evidence integral evaluations required by nested sampling and MCMC typically exhibits only mild scaling with the dimensionality of parameter space, as long as the integrand is well-behaved (i.e. unimodal, approximately Gaussian, with linear degeneracies between parameters)~\cite{RobertsandRosenthal, Skilling2006}.  However, likelihoods or priors of higher complexity with features such as multimodality, non-Gaussianity or non-linear degeneracies may lead to a significant increase in the number of function calls needed.

This is not problematic \textit{per se}, but may become an issue when coupled with evidence integrands that are expensive to calculate.  Here, the cost of a single evaluation depends both on the model and the observable under consideration.  To give an example from a cosmological context, considering \textit{Planck} Cosmic Microwave Background (CMB) anisotropy data~\cite{Planck:2019nip}, computation times for theory plus likelihood on a modern CPU may range from a fraction of a second for base $\Lambda$CDM cosmology to $\mathcal{O}(1~\mathrm{s})$ in extended models (e.g., $\Lambda$CDM with non-zero spatial curvature~\cite{Lewis:1999bs}, or inflationary features~\cite{Chluba:2015bqa})).  However, for other models or observables where costly non-perturbative simulations must be performed (e.g., N-body~\cite{Springel:2020plp}, hydrodynamics~\cite{Valentini:2025vhq}, lattice~\cite{Figueroa_2021} or numerical relativity~\cite{Aurrekoetxea:2024ypv}), it might take hours or even days.  In the future, as the precision of observations improves, there will be a trend for the corresponding likelihoods to become more elaborate, necessitating at the same time more accurate (and therefore more expensive) theory calculations, thereby exacerbating the issue.\footnote{In addition, new data may require the introduction of additional nuisance parameters, causing the dimensionality of the problem to increase as well.}

A possible solution to this problem lies in replacing the expensive evidence integrand with a fast approximation (a.k.a.\ emulator or surrogate) that is trained on the real integrand – and then performing the integration as usual, e.g. with nested sampling, on the emulated integrand.  

There are many examples of cosmological theory emulators for CMB or matter power spectra, see for instance Refs.~\cite{Jimenez:2004ct,Fendt:2006uh, Lawrence_2010,Winther:2019mus,Albers:2019rzt,Euclid:2020rfv,Moran_2022,Gunther:2022pto,SpurioMancini2022,Piras:2023aub,Gunther:2023xhh,Gunther:2025xrq}.  Pure theory emulators have the advantage of being applicable to different (combinations of) data sets without having to retrain, though their construction is of course specific to a particular kind of observable, and they may only be of limited use in cases where the computation of the likelihood function itself is expensive.  In this work, we opt to emulate the full likelihood\footnote{more precisely, the logarithm of the unnormalised posterior} instead.  While such an approach does require separate emulators to be trained for each combination of model and data set under consideration, it also makes the method of emulator construction agnostic to whatever observable the likelihood implicitly depends on.  The applicability of our methodology is therefore not limited to problems involving, say, CMB or matter power spectrum data, or cosmology for that matter, but it can be used on any function that is positive definite and integrable over the support of the prior.

Starting from the assumption of the evidence integrand being expensive to compute, our task is therefore to build an emulator that satisfies the following criteria:
\begin{itemize}
    \item[1)]{It is sufficiently precise, i.e. the total evidence uncertainty $\sigma_{\ln \mathcal{Z}}$ should be dominated by the integration uncertainty, not the emulator uncertainty,}
    \item[2)]{Its construction should endeavour to minimise the number of likelihood computations required to achieve criterion 1), and}
    \item[3)]{It does not make any prior assumptions about the shape of the likelihood function and is thus applicable to arbitrary likelihoods.}
\end{itemize}
The second criterion disfavours static pre-determined training strategies such as grid scans, Latin hypercube~\cite{Loh1996, Santner2018, Rogers2019} or hypersphere~\cite{Nygaard:2024lna} sampling or on-the-fly emulation methods with underlying random walk/sampling~\cite{Aslanyan:2015vfa,Gunther:2023xhh,Gunther:2025xrq} in favour of training via an active sampling strategy.  In other words, we want to employ an algorithm to dynamically choose where to evaluate the next sample, and iteratively ``learn'' the target function as more samples are collected.

Our guiding principle in designing such an algorithm will be the question:
\textit{Given what we know about $\hat{\mathcal{P}}$ from the samples already taken, where should it be evaluated next such that we maximise the expected reduction in the uncertainty of $\ln \mathcal{Z}$?}
As we shall discuss in more detail below, an emulator based on Gaussian Process Regression (GPR)~\cite{RasmussenWilliams2006} combined with a Bayesian Optimisation (BO)-inspired~\cite{Shahriari2016, Frazier2018,garnett_bayesoptbook_2023} active acquisition strategy ideally suits our requirements\footnote{See also Ref.~\cite{Janken:2025wlq} for a recent example of a neural network-based likelihood emulator.}, as it can be used not only to address this question, but in addition provide us with a reliable convergence diagnostic, i.e., the ability to detect when the emulator meets criterion 1).  

GPR with BO has been applied to cosmology for finding likelihood extrema~\cite{Hamann:2021eyw} or speeding up parameter inference~\cite{Leclercq2018,Pellejero-Ibanez:2019enw,Takhtaganov:2019ywj,Gammal:2022eob,Torrado:2023cbj,Rocher:2023asn}. The framework presented in our work can perform these tasks as well, but its focus lies on facilitating evidence estimation and model selection for expensive likelihoods.  \rev{Closer in spirit to this goal are Refs.~\citep{osborne2012active} and \citep{Chai2019bayesianquadrature}.  Both of them employ Bayesian quadrature; the former for the purpose of evaluating the evidence integral, the latter for the related differential entropy integral.  While Bayesian quadrature is known to perform well in lower-dimensional cases with well-behaved integrands~\citep{hennig2022probabilistic}, it may not be the most suitable method for the $\mathcal{O}(20)$-dimensional problems typically encountered in cosmology due to unfavourable scaling with the number of dimensions. In contrast, our approach relies on well-established Monte Carlo-based sampling techniques (which enjoy more favourable scaling) to compute the evidence integral and its uncertainty, as well as the acquisition function.} 

\begin{figure}[ht]
    \centering
    \includegraphics[width=0.99\linewidth]{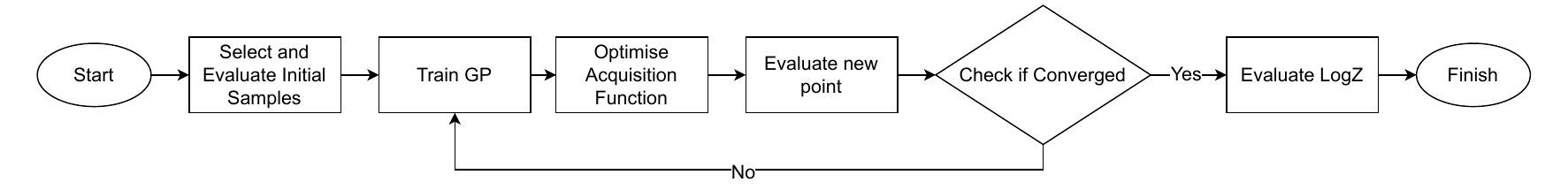}
    \caption{Flowchart of the general \texttt{BOBE} framework for constructing a likelihood emulator and computing the log-evidence.}
    \label{fig:BOBEFlowSimple}
\end{figure}

\rev{The main novel contribution of this work is the development of a BO-based active learning framework for constructing a likelihood emulator, aimed specifically at Bayesian evidence estimation and model selection – we thus name it \texttt{BOBE} (Bayesian Optimisation for Bayesian Evidence).
A basic flowchart of the \texttt{BOBE} algorithm is shown in Figure~\ref{fig:BOBEFlowSimple} and the \texttt{BOBE} \texttt{Python} package is available for download from \url{https://github.com/Ameek94/BOBE}.}

This paper is structured as follows: in Section~\ref{sec:GP}, we provide a brief review of Gaussian Process Regression, followed by a derivation of our acquisition function in Section~\ref{sec:WIPStd} and a description of the \texttt{BOBE} algorithm in Section~\ref{sec:bobe}.  A validation of \texttt{BOBE}, along with an analysis of its efficiency and results of its application to real cosmological problems is presented in Section~\ref{sec:results_and_validation} before we conclude in Section~\ref{sec:conclusion}. 

\section{Gaussian Process Regression
\label{sec:GP}}
We briefly recap the central concepts underlying Gaussian Process Regression here – for more details the reader is referred to the textbook by Rasmussen and Williams~\cite{RasmussenWilliams2006}.
GPR is a non-parametric regression method that models a real-valued target function $f(\boldsymbol{x}), \boldsymbol{x} \in \mathbb{R}^N$\footnote{In our case, we identify $\boldsymbol{x}$ with the model parameters $\theta$ and $N$ is the dimensionality of parameter space.} in terms of a Gaussian Process, i.e., one imposes a prior belief that $f$ is described by a collection of jointly Gaussian random variables with mean function $m(\boldsymbol{x})$ and whose covariance is determined by a kernel function $k(\boldsymbol{x},\boldsymbol{x}')$:
\begin{align}
    f(\boldsymbol{x}) \sim \mathcal{GP}(m(\boldsymbol{x}), k(\boldsymbol{x}, \boldsymbol{x^{\prime}})).
\end{align}
Given a training set $\mathcal{T}$ consisting of $n$ noisy ``measurements'' of the target function $f$ with iid noise variance $\sigma_\mathrm{n}^2$,
\begin{align}
    \mathcal{T} = \{ X, \boldsymbol{y} \} \quad
     X = [\boldsymbol{x}_1, \ldots, \boldsymbol{x}_n]^\top, \quad 
    \boldsymbol{y} = [f(\boldsymbol{x}_1), \ldots, f(\boldsymbol{x}_n)]^\top,
\end{align}
the GP prior implies that the training values $\boldsymbol{y}$ and the function value at a new test point $\boldsymbol{x}_{*}$ follow a joint normal distribution:
\begin{align}
\begin{bmatrix}
\boldsymbol{y} \\ f(\boldsymbol{x}_*)
\end{bmatrix}
\sim \mathcal{N}\!\left(
\begin{bmatrix}
\boldsymbol{m}(X) \\ m(\boldsymbol{x}_*)
\end{bmatrix},
\begin{bmatrix}
K(X,X) + \sigma_\mathrm{n}^2 \mathbb{1} & \boldsymbol{k}(X,\boldsymbol{x}_*) \\
 \boldsymbol{k}(X,\boldsymbol{x}_*)^{\top}  & k(\boldsymbol{x}_*,\boldsymbol{x}_*)
\end{bmatrix}
\right),
\end{align}
where $\{\boldsymbol{m}(X)\}_p = m(X_p)$, $\{K(X,X)\}_{pq} = k(\boldsymbol{x}_p, \boldsymbol{x}_q)$ and $\{\boldsymbol{k}(X,\boldsymbol{x}_*)\}_p = k(X_p, \boldsymbol{x}_*)$. This in turn implies that at a test point $\boldsymbol{x}_*$ the result of the regression is the posterior distribution
\begin{equation}
f(\boldsymbol{x}_*) |\mathcal{T} \sim \mathcal{N}(\mu (\boldsymbol{x}_*), \sigma^2(\boldsymbol{x}_*)),
\end{equation}
with the posterior mean $\mu(\boldsymbol{x}_*)$ and its variance $\sigma^{2}(\boldsymbol{x}_*)$ (which quantifies the uncertainty of the regression) given by
\begin{align}
\mu(\boldsymbol{x}_*) &= m(\boldsymbol{x}_*) + \boldsymbol{k}(X,\boldsymbol{x}_*)^{\top}  \left(K(X,X) + \sigma_\mathrm{n}^2 \mathbb{1}\right)^{-1} \big(\boldsymbol{y} - \boldsymbol{m}(X)\big), \\[6pt]
\sigma^2(\boldsymbol{x}_*) &= k(\boldsymbol{x}_*,\boldsymbol{x}_*) - \boldsymbol{k}(X,\boldsymbol{x}_*)^{\top}  \left(K(X,X) + \sigma_n^2 \mathbb{1}\right)^{-1} \boldsymbol{k}(X,\boldsymbol{x}_*).
\end{align}

\subsection{GP mean, kernel and hyperparameters}
Evidently, the above result depends on the choice of GP mean function $m(\boldsymbol{x})$ and kernel $k(\boldsymbol{x},\boldsymbol{x}')$.  By fixing the functional forms of $m$ and $k$, the problem can be reduced to choosing instead a finite number of hyperparameters $\phi$.

If we make the assumption that no part of parameter space is \textit{a priori} preferred over any other, then the natural choice for the GP mean becomes
\begin{equation}
    m(\boldsymbol{x}) = m_0 =\mathrm{const}.
\end{equation}
As for the kernel, many options have been explored in the literature~\cite{duvenaud_2014}.  One can generally expect likelihoods to be continuous, differentiable functions, in which case a Gaussian kernel is a suitable choice:
\begin{equation}
    k(\boldsymbol{x},\boldsymbol{x}') = \sigma_f^2 \, \exp \left[ -\tfrac{1}{2} \sum_{i=1}^N \left( \frac{x_i - x_i'}{\ell_i}\right)^2\right] + \sigma_n^2 \, \delta(\boldsymbol{x} - \boldsymbol{x}').
\end{equation}
Besides the already mentioned noise variance $\sigma_\mathrm{n}^2$, this kernel introduces $N+1$ additional hyperparameters: the outputscale $\sigma_f^2$ which determines the width of the GP prior, and one correlation length parameter $\ell_i$ for each of the $N$ dimensions of parameter space.  Notably, this kernel depends only on distances between points in parameter space, not on location, consistent with our previous assumption that no part of parameter space is \textit{a priori} special.

The quality of a GPR's fit to the training data can be quantified in terms of the marginal log-likelihood (MLL):
\begin{equation}
\label{eq:mll}
    \begin{split}
    \ln p(\mathbf{y}|X, \phi) =& -\tfrac{1}{2} \,(\mathbf{y} - \boldsymbol{m}(X))^{\top} \left(K(X,X) + \sigma^{2}_\mathrm{n}\mathbb{1}\right)^{-1}(\mathbf{y} - \boldsymbol{m}(X)) \\
    & - \tfrac{1}{2} \ln \left| K(X,X) + \sigma^{2}_\mathrm{n} \mathbb{1} \right| - \tfrac{n}{2}\ln 2\pi.
    \end{split}
\end{equation}
Applying Bayes' theorem and multiplying the MLL with hyperparameter prior probabilities $p(\phi)$, one obtains a posterior probability density for the hyperparameters given the training data:
\begin{equation}
    \label{eq:mlp}
    \ln p(\phi|X,\boldsymbol{y}) = \ln p(\phi|\mathcal{T}) = \ln p(\mathbf{y}|X, \phi) + \ln p(\phi) + \mathrm{const}.
\end{equation}
By maximising this quantity we can thus let the data decide which values of the hyperparameters are most appropriate.\footnote{In a fully Bayesian treatment, one could marginalise $p(\phi|\mathcal{T})$ over $\phi$ instead~\cite{Lalchand2020FullyBayesianGPR, Lalchand2022SparseGPHyper}, but in practice we found that for our problem this yields very little gain at the cost of substantial additional computational overhead.}  A summary of our hyperparameters and  their treatment can be found in Tab.~\ref{tab:hyper}.

\begin{table}
    \centering
    \begin{tabular}{lccc}
    \hline
    Hyperparameter & Symbol & Value & Bounds  \\
    \hline
    Outputscale & $\sigma_f^2$ & * & $[10^{-4},10^{8}]$ \\
    Correlation scale for $\theta_i$ & $\ell_i$ & * & $[0.01, 10]$ \\    
    GP mean & $m_0$ & $\frac{1}{n} \sum_{i=1}^n f(\boldsymbol{x}_i)$ & n/a \\   
    Noise variance & $\sigma_\mathrm{n}^2$ & $10^{-12}$ & n/a \\ 
    \hline
    \end{tabular}
    \caption{List of GP hyperparameters.  An asterisk denotes hyperparameters whose values are determined by maximising their marginalised log posterior~Equation~\eqref{eq:mlp}. Though likelihood evaluations are not usually subject to noise, a small non-zero noise variance improves numerical stability. The upper and lower bounds on outputscale and correlation scale are given in terms of standardised input, see Sec.~\ref{sec:bobeinit} for details.
    \label{tab:hyper}}
\end{table}

\subsection{Emulation}
Which physical quantity should the target function $f$ be identified with for our purpose?  At first glance, one might na{\"i}vely pick the integrand of the evidence integral, $\hat{\mathcal{P}}(\theta|\mathcal{M})$.   
However, there are two major advantages to emulating its logarithm, $\ln \hat{\mathcal{P}}(\theta|\mathcal{M})$,
instead~\cite{osborne2012active}.

Firstly, this choice ensures that the emulator does not return unphysical negative values for the posterior.
Secondly, a given absolute error in $\ln \hat{\mathcal{P}}$ becomes a relative error in $\hat{\mathcal{P}}$ itself, and the corresponding absolute error of $\hat{\mathcal{P}}$ (which is what is relevant for the evidence uncertainty) scales with $\hat{\mathcal{P}}$, i.e., far in the tails of $\hat{\mathcal{P}}$ it will not be necessary for the emulation to be extremely precise.

We do note that $\ln \hat{\mathcal{P}}$ is in principle unbounded from below, and having to cover too large a dynamical range can affect the accuracy of the emulation.  However, noting that if $\hat{\mathcal{P}}$ is small enough it becomes essentially indistinguishable from zero, this issue can be addressed by introducing an effective floor by means of a classifier, see Sec.~\ref{sec:classifier} for more details.

\section{Bayesian Optimisation and the Weighed Integrated log-Posterior Standard Deviation (WIPStd)
\label{sec:WIPStd}}
As seen above, GPR gives us not only an estimate of the expectation value of the target function, but also an uncertainty, opening up the opportunity to use our knowledge from the GPR to inform an active exploration strategy, in contrast to methods that by construction rely on random sampling or random walks.

Let us designate the GPR output for a training set $\mathcal{T}_t$ of $t$ input data points by $(\mu_t(\boldsymbol{x}), \sigma_t(\boldsymbol{x}))$. In Bayesian Optimisation~\cite{Shahriari2016,Frazier2018,garnett_bayesoptbook_2023}, the next point at which to evaluate the target function, $\boldsymbol{x}_{t+1}$, is determined by extremising an acquisition function 
%$\mathcal{A}(\mu_t(\boldsymbol{x}), \sigma_t(\boldsymbol{x}),\boldsymbol{x})$.  
$\mathcal{A}(\boldsymbol{x})$. The choice of acquisition function depends on the problem at hand; common examples include Expected Improvement (EI)~\cite{garnett_bayesoptbook_2023} for finding the optimum of the target function, or Integrated Posterior Variance (IPV)~\cite{Gorodetsky2016, ChemRxiv2025}/Integrated Mean Squared Prediction Error (IMSPE)\cite{Sacks1989, Crary2015, Crary2016} if one aims to maximise reduction in the uncertainty of the target function regression over all of parameter space.

Neither of these are quite ideally suited to our goal of maximising the decrease in the uncertainty of the evidence (i.e., an integral over a nonlinear mapping of the target function).  We will therefore proceed to construct our own acquisition function, starting from the estimate of the evidence at iteration $t$:

\begin{align}
    \zc_t = \int_{V} \D \boldsymbol{x} \, e^{\mu_t(\boldsymbol{x})}
\end{align}
where $V$ is the domain of prior support.  As a rough but conservative estimate, we quantify the effect of $\sigma_t(\boldsymbol{x})$ on $\zc_t$ by 
\begin{align}
    \zc_{t, \pm} = \int_{V} \D \boldsymbol{x} \, e^{\mu_t(\boldsymbol{x}) \pm \sigma_t(\boldsymbol{x})}.
\end{align}
Linearising the exponential for $\sigma_t(x) \ll 1$ 
\begin{align}
     \zc_{t, \pm} &\approx \int_{V} \D \boldsymbol{x} \, e^{\mu_t(\boldsymbol{x})}(1 \pm \sigma_t(\boldsymbol{x}))\,
\end{align}
and introducing the shorthand
\begin{align}
    S_t \equiv \int_{V} \D \boldsymbol{x} \, e^{\mu_t(\boldsymbol{x})}\sigma_t(\boldsymbol{x}),
\end{align}
we have at first order in $\sigma$
\begin{align}
    \zc_{t, \pm} \approx \zc_{t} \pm S_t, \qquad 
    \ln \zc_{t, \pm} \approx \ln \left( \zc_{t} \pm S_t \right).
\end{align}
For $S_t \ll \zc_t$ we Taylor-expand the logarithm around $S_t = 0$,
\begin{align}
    \ln \left( \zc_{t} \pm S_t \right) \approx \ln \zc_t \pm \frac{S_t}{\zc_t}.
\end{align}
The symmetric uncertainty in the log-evidence is then approximately
\begin{align}
    \sigma_{\ln \mathcal{Z},t} &\approx \frac{1}{2}\left[\ln \zc_{t, +} - \ln \zc_{t, -} \right] \\
    &\approx \frac{1}{2}\left[\ln \zc_t + \frac{S_t}{\zc_t} - \ln \zc_t + \frac{S_t}{\zc_t} \right] =  \frac{S_t}{\zc_t}\\
    &=  \int_{V} \D \boldsymbol{x} \, \frac{ e^{\mu_t(\boldsymbol{x})}}{\zc_t} \sigma_t(\boldsymbol{x}).
\end{align}
The factor  ${e^{\mu_t(\boldsymbol{x})}}/{\zc_t}$
in the integrand is the GP expectation of the (normalised) posterior probability density $\mathcal{P}_t(\boldsymbol{x})$. So the uncertainty in $\ln \zc_t$ becomes an integral over the GP uncertainty, weighted by the posterior:
\begin{align}
    \sigma_{\ln \mathcal{Z},t}\approx \int_{V} \D \boldsymbol{x} \, \mathcal{P}_t(\boldsymbol{x}) \sigma_t(\boldsymbol{x}).
\end{align}
This explicitly demonstrates the advantage of emulating $\ln \hat{\mathcal{P}}$: we do not require GPR precision to be the same everywhere in parameter space, high precision is only required in regions of high posterior probability.

If we were to augment our training set by adding a new ``fantasy'' data point $(\boldsymbol{x}', 
\mu_t(\boldsymbol{x}'))$ (i.e., rather than evaluating $f(\boldsymbol{x}')$, we pretend the function value at $\boldsymbol{x}'$ is the GPR expectation), and updated the GPR (while keeping the hyperparameters the same), the augmented GP's standard deviation can be calculated via \cite{RasmussenWilliams2006}
\begin{align}
    \sigma^{2}_{t+1}(\boldsymbol{x} | \boldsymbol{x}') = \sigma^2_{t}(\boldsymbol{x}) - \frac{\mathrm{cov}(f(\boldsymbol{x}), f(\boldsymbol{x}'))^2}{\sigma^{2}_t(\boldsymbol{x}')},
\label{eq:varupdate}
\end{align}
where
\begin{align}
    \mathrm{cov}_t(f(\boldsymbol{x}), f(\boldsymbol{x}')) =k(\boldsymbol{x, x'}) - \boldsymbol{k}(X,\boldsymbol{x})^{\top}\left(K(X,X) + \sigma_\mathrm{n}^{2} \mathbb{1}\right)^{-1}\boldsymbol{k}(X,\boldsymbol{x'}).
\end{align}
The resulting evidence uncertainty upon the addition of the fantasy data point is a function of $\boldsymbol{x}'$: 
\begin{align}
     \sigma_{\ln \mathcal{Z},t}(\boldsymbol{x'}) \approx \int_{V} \D \boldsymbol{x} \, \mathcal{P}_t(\boldsymbol{x})\sigma_{t+1}(\boldsymbol{x} | \boldsymbol{x}')\,.
     \label{eq:WIPStd_analytic}
\end{align}
We dub the acquisition function defined by Equation~\eqref{eq:WIPStd_analytic} ``Weighted Integrated log-Posterior Standard Deviation'' (WIPStd).

In practice, we approximate the WIPStd integral using Monte Carlo (MC) sampling, by drawing $M$ samples $\{ \boldsymbol{x}_j \}_{j=1}^{M}$ from the posterior $\mathcal{P}_t$:
\begin{align}
    \sigma_{\ln \mathcal{Z},t}(\boldsymbol{x'}) \approx \widehat{\mathfrak{W}}(\boldsymbol{x}') \equiv \frac{1}{M} \sum_{j=1}^{M} \sigma_{t+1}(\boldsymbol{x}_j | \boldsymbol{x}')
\label{eq:WIPSTd_Approx}
\end{align}
and adopt $\widehat{\mathfrak{W}}(\boldsymbol{x}')$ as our acquisition function, i.e., the next point to evaluate the target function at is chosen as
\begin{align}
    \boldsymbol{x}_{t+1} = \displaystyle \arg \min_{\boldsymbol{x}'} \widehat{\mathfrak{W}}(\boldsymbol{x}'). 
\end{align}
As only one new observation is being added at a time, we can utilise a rank-one update of the Cholesky factorisation of the kernel matrix, requiring only $\mathcal{O}(N^2)$ operations~\cite{garnett_bayesoptbook_2023}.  Thus, the cost of acquisition function optimisation remains sub-dominant to that of GP training which goes as $\mathcal{O}(N^3)$~\cite{RasmussenWilliams2006}.

Equation~\eqref{eq:WIPSTd_Approx} highlights another extremely convenient property of this construction, namely that the acquisition function $\widehat{\mathfrak{W}}$ is a direct measure of the \red{emulator's} log-evidence uncertainty $\sigma_{\ln \mathcal{Z}}$ \rev{which in turn is a useful proxy for the error (i.e., the difference from the true value) in the evidence estimate.}\footnote{\rev{As the GP converges to the target function, not only do the linear approximations in deriving $\mathfrak{W}$ from the evidence uncertainty become better, but $\sigma_{\ln \zc}$ will also become a more reliable estimate of the error in the evidence estimate.}} This means it can also be used to track how well the emulator has converged to the target function and thus help us decide when it is precise enough for our purpose, i.e., when further acquisition of samples becomes unnecessary and the BO loop can be stopped.

\section{The \texttt{BOBE} algorithm \label{sec:bobe}}
In this Section, we provide a more technical discussion of the central ingredients of \texttt{BOBE} and present a pseudocode schematic in Algorithm~\ref{alg:gp_wipstd}. 

\subsection{Initialisation \label{sec:bobeinit}}
\paragraph{Initial training data:} It is not strictly necessary for the algorithm to begin with initial samples of the target function; a GPR without data simply returns the GPR prior.
But it is of course also possible, and sometimes advantageous, to start from a non-empty initial training set, for instance if samples from previous runs for the same prior/model/data  combination\footnote{or postprocessed samples for different priors or data} are available.  Or in case one has prior knowledge about characteristics of the target function that are not already represented in the parameter priors, e.g., the (approximate) location of the global posterior maximum, or properties like suspected multimodality.  If the GP already knows where the interesting regions of the target function are rather than having to discover them first, the algorithm will tend to converge more quickly and be less likely to suffer from premature convergence.

In addition to permitting starts from a user-supplied training set, \texttt{BOBE} includes options to
build an initial training set by generating
\begin{itemize}
    \item{random samples from the prior or an arbitrary user-defined reference distribution, or}
    \item{quasi random, low discrepancy Sobol sequence~\cite{sobol1967distribution} samples over the prior support (see Ref.~\cite{eriksson2021highdimensionalbayesianoptimizationsparse}).}
\end{itemize}
Either of these two options essentially enforces an initial wide exploration of parameter space before active sampling commences.  Having too many samples initially generated this way may negatively impact overall efficiency though – we recommend that one should aim to keep the number of initial points below $10$--$20\%$ of the total likelihood evaluation budget.

\paragraph{Input standardisation:} In order to improve numerical stability, we initially map the prior volume of parameter space to the unit hypercube.  In addition, we standardise the target function values such that the training points have zero mean and unit variance – this is repeated every time the training set is updated.  Internally, the code therefore only deals with standardised input.  All quantities are mapped back to their physical values in the final output.

\subsection{GP Training}
Starting with the initial training set, and every time the training set is expanded, we perform a GPR as outlined in Section~\ref{sec:GP}.  The output and correlation scale hyperparameters are determined by applying a gradient-based optimiser (\red{by default this is performed by the L-FBGS-B algorithm} included in \texttt{scipy}~\cite{2020Scipy}) to find the values that maximise the MLL.  The optimiser is restarted multiple times from random initial points in hyperparameter space, plus one restart from the previous iteration's optimal values. Hyperparameters typically tend to stabilise once the training set becomes large enough so it may not be necessary to optimise them (a task that scales like $N^3$ for a GP with $N$ training points) every single time a sample is added. \red{Within \texttt{BOBE} this is controlled by the parameter $\tau_{\mathrm{refit}}$ which ensures the GP is only retrained once at least $\tau_{\mathrm{refit}}$  new points have been added to the GP training set}.

\subsection{Acquisition}
\begin{figure}
    \centering
    \includegraphics[width=0.95\linewidth]{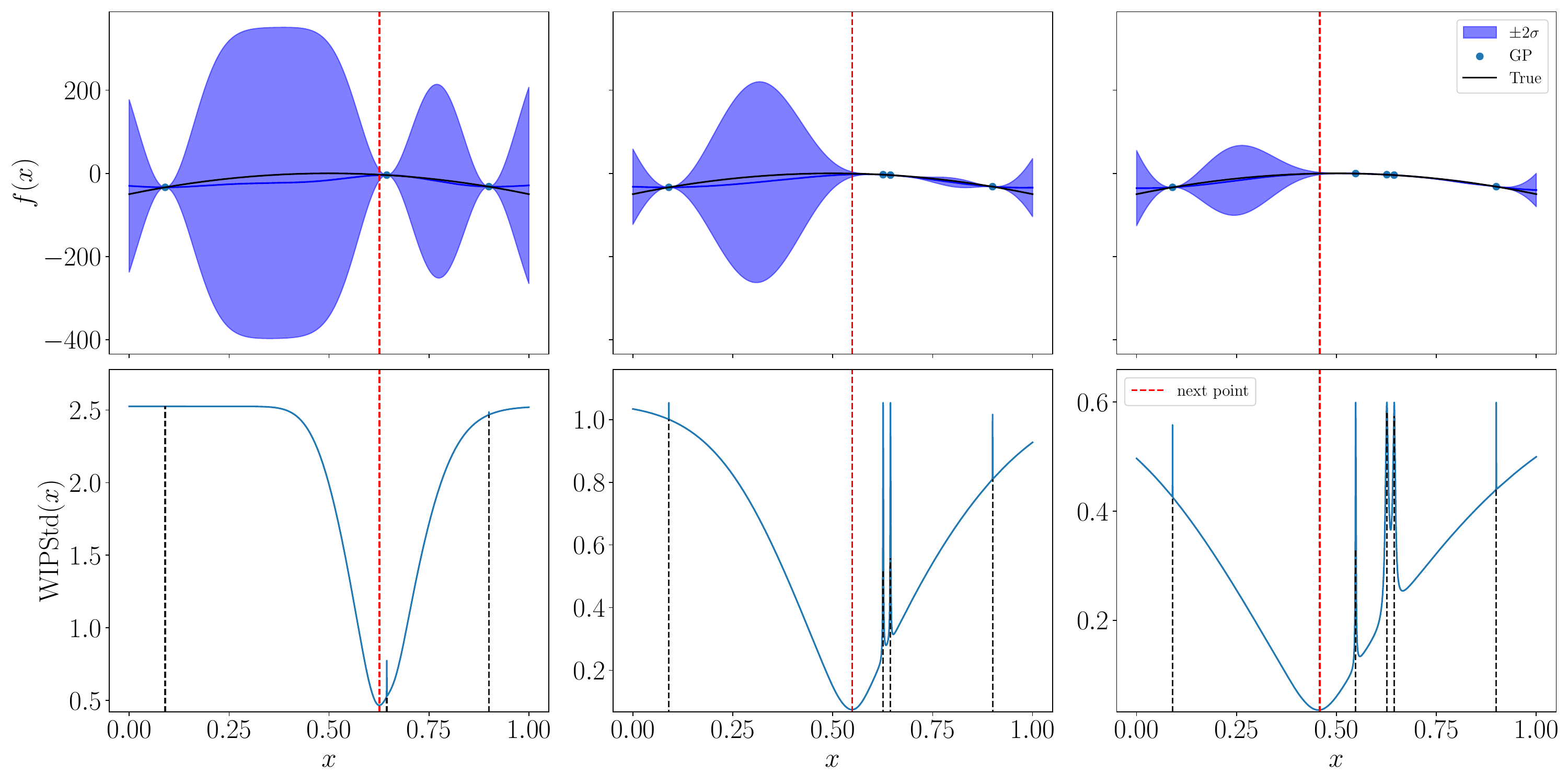}
    \caption{Demonstration of the WIPStd acquisition function in action across three consecutive iterations. \textbf{Top row:} objective function (black solid line) and GP mean prediction (blue points) with evaluation locations (blue circles) and $\pm 2 \sigma(x)$ GP uncertainty region (blue bands). \textbf{Bottom Row:} $\mathrm{WIPStd}(x)$, approximated by $\widehat{\mathfrak{W}}(x)$ (light blue solid line) with evaluation locations (grey dashed lines) and location of next evaluation (red dashed line). Note the rapid decrease in $\mathrm{WIPStd}(x)$ as additional samples are added}.
    \label{fig:wipstd_plot}
\end{figure}

\paragraph{Evaluating and optimising WIPStd:} To evaluate the acquisition function $\widehat{\mathfrak{W}}$, we first generate a set of samples from the posterior\red{, $n_\mathfrak{W}$}\footnote{When choosing the number of samples, one should aim to have sufficient samples to properly represent the emulated posterior without introducing unnecessary computational overhead in the calculation of the acquisition function. Generally, higher dimensional and more complex target functions require more samples.}, utilising either \texttt{NumPyro}'s~\cite{numpyro} No-U-Turn (NUTS) Hamiltonian Monte Carlo sampler~\cite{hoffman2011nouturnsampleradaptivelysetting} or the \texttt{Dynesty}~\cite{Speagle2020} nested sampler.  $\widehat{\mathfrak{W}}$ as a function of the location of a fantasy observation can then be computed using Equation~\eqref{eq:WIPSTd_Approx}.
A gradient-based optimiser is then used to find the point that minimises the acquisition function \red{and at which the target function will be evaluated next. As in the hyperparameter optimisation step above, our default choice of optimiser is the L-FBGS-B algorithm as implemented in \texttt{scipy}~\cite{2020Scipy}.}
We illustrate this process for a simple 1-dimensional example in Figure~\ref{fig:wipstd_plot}.  As we can see in the lower panels, the structure of the acquisition function can become quite complicated as more samples are taken, and it is not unlikely for the optimiser to only discover a local minimum and miss the global one.  However, in the long run imperfect optimisation at this step will not pose a major issue for the performance of the algorithm.  Eliminating significant local minima of the acquisition function makes discovery of the global minimum more likely in the following iterations, so in most cases this will only affect the order in which points are sampled.

\paragraph{Batch acquisition:} Rather than immediately computing the target function after one round of acquisition minimisation, and adding the new data point to the training set, one can also (temporarily) add a fantasy observation at the minimum to the training set and repeat the whole acquisition process, multiple times if desired.  This method is also known as the Kriging believer method \cite{KrigingVarianceBarnes1992, Ginsbourger2010, chevalier2012correctedkrigingupdateformulae, Chevalier:978-3-642-44973-4,JMLR:v15:desautels14a,ginsbourger:hal-00260579,Torrado:2023cbj} and allows the selection of a whole batch of candidate points for target function evaluation in one go – which has the big advantage of letting us perform expensive target function computation in parallel,  \rev{though in batch acquisition one does not necessarily select the optimal sample beyond the first one of each batch, so its usage may incur a potential loss to the overall efficiency of the algorithm}.

Once a batch is completed, the minimum value of the acquisition function for the final candidate point is retained to be used in the convergence test step, the temporary fantasy points are removed from the training set, and the true $\ln \hat{\mathcal{P}}$ is computed for all candidate points.

\subsection{Classifier filtering \label{sec:classifier}}
In many realistic applications, especially those with higher-dimensional parameter spaces, it is not atypical for the evidence integral to be dominated by the contribution of a small fraction of the prior volume, and for most of the prior volume to be occupied by the tails of the posterior which contain a negligible part of the total evidence. It is therefore unnecessary to aim for a very precise emulation of $\ln \hat{\mathcal{P}}$ in the tails. In fact, it may even be counterproductive to do so, since increasing the range of $\ln \hat{\mathcal{P}}$ one tries to emulate may negatively affect the ability of the GPR to precisely model the important high-posterior regions.

This issue can be avoided by introducing an effective floor for $\ln \hat{\mathcal{P}}$ by means of a classifier that distinguishes between relevant and irrelevant parts of parameter space – an approach that was also used in \texttt{Gpry}~\cite{Gammal:2022eob}.  
Data points for which $\ln \hat{\mathcal{P}}_\mathrm{max} - \ln \hat{\mathcal{P}}$ exceeds a threshold $\Delta$ (selected to be large enough such that $\hat{\mathcal{P}}_\mathrm{max} \,e^{-  \Delta}$ is functionally indistinguishable from $\hat{\mathcal{P}} = 0$) will not be passed to the GP, and in regions classified as irrelevant, the emulator will set $\ln\hat{\mathcal{P}}$ to the effective floor value $\ln \hat{\mathcal{P}}_\mathrm{max} - \Delta$.  Details of our implementation are provided in Appendix~\ref{sec:CLF}.

An additional benefit of using a classifier is that it makes the emulator immune to catastrophic failure in cases where the likelihood function should fail and return unphysical values for the log-likelihood (e.g., \texttt{NaN} or $- \infty$) – in this case, the corresponding part of parameter space would be  classified as irrelevant, $\ln \hat{\mathcal{P}}$ set to the effective floor value and it would not contribute to the evidence.

The classifier is retrained each time the training set is updated. Importantly, no data are discarded in this process; sub-threshold points are simply not given to the GP.
However, let us add a cautionary note here: in case of multimodal posteriors, activating the classifier too soon may preclude the discovery of isolated local maxima.

\subsection{Convergence}
As pointed out in Section~\ref{sec:WIPStd}, WIPStd (and $\widehat{\mathfrak{W}}$) are specifically constructed to approximate the emulator uncertainty $\sigma_{\ln \mathcal{Z}}$, i.e., we can use our acquisition function directly to track the emulator's convergence.  Reliable model comparison demanding an overall log-evidence accuracy of $\mathcal{O}(0.1)$, and the typical integrator accuracy being $\sim 0.1$ on the log-evidence for numerical integration via nested sampling, suggest that a convergence threshold of \mbox{$\widehat{\mathfrak{W}} < \mathcal{O}(0.1)$} should be a reasonable convergence criterion for most applications. 
\rev{In problems with complex target functions, and particularly for multimodal posteriors, there is the possibility of \texttt{BOBE} declaring convergence too early, e.g., before all relevant modes are discovered by the emulator.  This limitation is of course not unique to \texttt{BOBE}, and similarly applies to other active or stochastic sampling methods. We will discuss some practical mitigation strategies in the context of a multimodal example in more detail in Section~\ref{sec:results_and_validation}, but let us point out for now that the risk of possible premature convergence due to a random downward fluctuation in $\widehat{\mathfrak{W}}$ can be decreased by demanding that $\widehat{\mathfrak{W}}$ be smaller than the convergence threshold for $n_\mathrm{iters}$ consecutive iterations/batches rather than terminating the first time the threshold is crossed.}

Once the convergence criterion is met, the samples selected in the last iteration are evaluated and added to the training set, the classifier (if used) is retrained and GPR performed one final time. The emulator is now complete and ready to use.

We then evaluate the Bayesian evidence of the model with a high precision nested sampling run of \texttt{Dynesty}~\cite{Speagle2020} on the emulator.\footnote{If desired for diagnostic reasons, it is of course also possible to compute an evidence estimate with an iteration of the emulator that does not yet meet the convergence criterion.}  While the \texttt{BOBE} emulator is designed for evidence computations, it can just as well be used for parameter inference, conveniently killing two birds with one stone.

\begin{algorithm}
\caption{Bayesian Optimisation for Bayesian Evidence (\texttt{BOBE}) }
\label{alg:gp_wipstd}
\begin{algorithmic}[1]
\Require Objective function $f(\boldsymbol{x})$, Precision $\varepsilon > 0$, Batch size $B$, $N_{\rm init}$ initial samples
\State Initialise dataset $\mathcal{T}_{\mathrm{train}} \gets \mathcal{T}_{\mathrm{init}} = \{\, (x_{i}^{\mathrm{(init)}}, y_{i}^{\mathrm{(init)}})\,\}_{i=1}^{N_{\mathrm{init}}}$

\If{Classifier Enabled}
    \State set classifier threshold $\gamma$
    \State Filter $\mathcal{T}_{\mathrm{train}}: \displaystyle \mathcal{T}_{\mathrm{train}} \gets \{(x_i, y_i) \in \mathcal{T}_{\mathrm{train}} : y_i > \max_{\mathrm{j}} y_{\mathrm{j}} - \gamma \}$
\EndIf

\State Optimise GP hyperparameters $\displaystyle \phi^\star \gets \arg \max_{\phi} \ln p(\mathbf{y}_{\mathrm{train}} | X_{\mathrm{train}}, \phi)$ (MLL)

\State Update GP model $f_\mathrm{GP}(x) | \mathcal{T}_\mathrm{train} \sim \mathcal{GP}\left(m(x), k(x, x^{\prime}, \theta^\star)\right)$
\While{$\sigma_{\ln \mathcal{Z}} \ge \varepsilon$}
    \State Draw posterior samples $\{ f_{\mathrm{GP}}^{(s)}(x) \}_{s=1}^S$ from $f_\mathrm{GP}(x)$
    \State Initialise batch $\mathcal{B} \gets \emptyset$ and temporary dataset $\mathcal{T}^{(1)}_{\mathrm{train}} \gets \mathcal{T}_{\mathrm{train}}$
    
    \For{$b = 1$ to B}
        \State Select next batch point:  $ \displaystyle \boldsymbol{x}_b = \arg \min_{\boldsymbol{x}} \widehat{\mathfrak{W}}^{(b)}(\boldsymbol{x})$ 
    
        \State Generate a fantasy observation: $\tilde{y}_b \sim f_{\mathrm{GP}}^{(b)}(\boldsymbol{x}_b)$
    
        \State Temporarily augment: $\mathcal{T}^{(b+1)}_{\mathrm{train}} \gets \mathcal{T}_{\mathrm{train}}^{(b)} \cup \{ \boldsymbol{x}_b,\, \tilde{y}_b \}$

        \State Update temporary GP: $f^{(b+1)}_\mathrm{GP}(\boldsymbol{x}) | \mathcal{T}^{(b+1)}_\mathrm{train}, \quad \mathcal{B} \gets \mathcal{B} \cup \{\boldsymbol{x}_b\}$

    \EndFor
    \State $\sigma_{\ln \mathcal{Z}} \gets \widehat{\mathfrak{W}}^{(B)}(\boldsymbol{x}_{B})$
    
    \State Evaluate $y_b = f(\boldsymbol{x}_b)$ for all $\boldsymbol{x}_b \in \mathcal{B}$
    
    \State Augment the training set: $\mathcal{T}_{\mathrm{train}} \gets \mathcal{T}_{\mathrm{train}} \cup \{ (\boldsymbol{x}_b, y_b) \}_{b=1}^{B}$

    \If{Classifier Enabled}
        \State \hspace{2em} Filter $\mathcal{T}_{\mathrm{train}}: \displaystyle \mathcal{T}_{\mathrm{train}} \gets \{(\boldsymbol{x}_i, y_i) \in \mathcal{T}_{\mathrm{train}} : y_i > \max_{j} y_{j} - \gamma \}$
    \EndIf
    
   \State Optimise GP hyperparameters $\displaystyle \phi^\star \gets \arg \max_{\phi} \ln p(\mathbf{y}_{\mathrm{train}} | \boldsymbol{X}_{\mathrm{train}}, \phi)$ (MLL)
    \State Update GP model $f_\mathrm{GP}(\boldsymbol{x}) | \mathcal{T}_\mathrm{train} \sim \mathcal{GP}\left(m(x), k(\boldsymbol{x}, \boldsymbol{x}^{\prime}, \theta^\star)\right)$
\EndWhile
\State $\ln \mathcal{Z}_\mathrm{GP} \gets \mathrm{NestedSample(f_\mathrm{GP}, \pi})$
\end{algorithmic}
\end{algorithm}

\section{Results and Validation}
\label{sec:results_and_validation}
\subsection{2-dimensional test functions}
To demonstrate the performance of our algorithm, we first consider a set of standard 2-dimensional test functions~\cite{Opt_benchmarks}. 
While the rather benign 2-dimensional Gaussian serves as a first proof of concept, the Gaussian Ring, Himmelblau's and Eggbox functions introduce various pathologies designed to challenge sampling algorithms.
Note that in all the following examples, we take these functions to describe the unnormalised posterior $\hat{\mathcal{P}}$, but GPR is performed on $\ln \hat{\mathcal{P}}$.

\noindent Our test functions are defined as follows:

\paragraph{2-dimensional Gaussian:}  
\begin{align}
    \ln \hat{\mathcal{P}}_{\mathrm{G-2D}}(x_1,x_2) = -\frac{1}{{2\sigma^2}}{\displaystyle \sum_{n=1}^{2} (x_n - \mu)^2}
\end{align}
where $x_1, x_2 \in \{ 0, 1\}$ and $\mu=0.5$, $\sigma=0.1$. 

\paragraph{Gaussian Ring:}
\begin{align}
    \ln \hat{\mathcal{P}}_{\mathrm{G-R}}(x_1, x_2) = -\frac{1}{2} \, \left(\frac{\sqrt{(x_1 - 0.5)^2 + (x_2 - 0.5)^2} - 0.2}{0.02}\right)^2 ,
\end{align}
where $x_1, x_2 \in \{ 0, 1\}$.

\paragraph{Himmelblau's Function:} 
\begin{align}
    \ln \hat{\mathcal{P}}_{\mathrm{H}}(x_1, x_2) = -\frac{1}{2} \, \left(0.1(x_1 + x_2^2 - 7)^2 + (x_1^2 + x_2 - 11)^2)\right),
\end{align}
where $x_1, x_2 \in \{ -4, 4\}$.

\paragraph{Eggbox:} 
\begin{align}
    \ln \hat{\mathcal{P}}_{\mathrm{E}}(x_1, x_2) = \left(2+\cos(4\pi x_1)\cos(4\pi x_2)\right)^3 - 27 ,
\end{align}
where $x_1, x_2  \in \{ 0, 1\}$. 

\red{These functions are defined such that the maximum of $\ln \hat{\mathcal{P}}$ is 0, corresponding to a maximum $\hat{\mathcal{P}}$ of 1.} 
\begin{figure}
    \centering
    \includegraphics[width=0.95\linewidth]{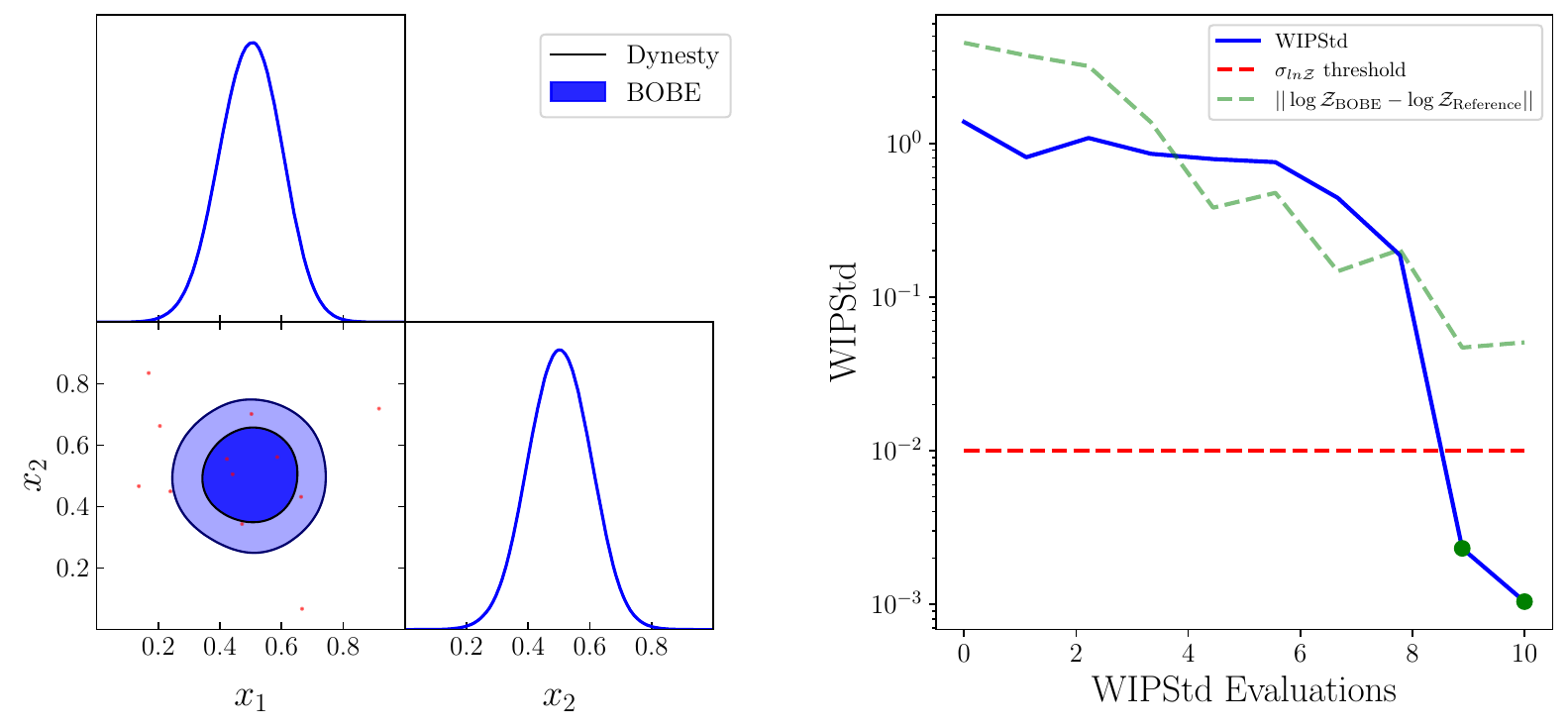}
    \includegraphics[width=0.95\linewidth]{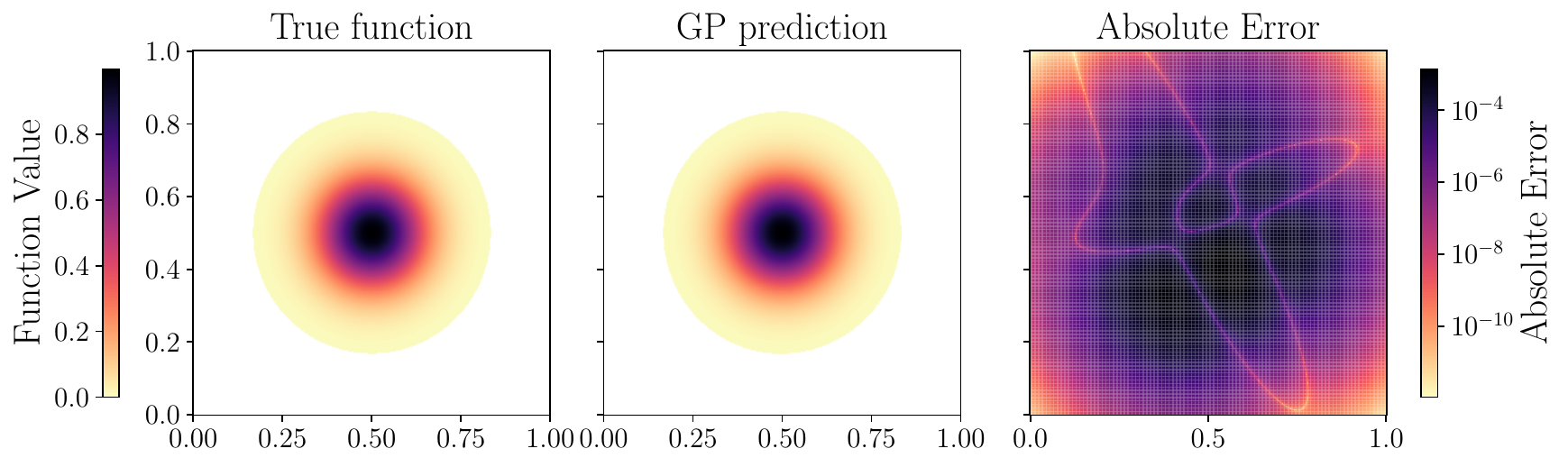}
    \caption{\texttt{BOBE} applied to a 2D Gaussian posterior. \textbf{Top Left}: Triangle plot comparing 1D and 2D posteriors contours obtained from the \texttt{BOBE} emulator (blue blobs/lines) and the true posterior (black lines). Locations of the training samples are indicated by red dots. \textbf{Top Right}: WIPStd convergence diagnostic (solid blue line) compared with the absolute difference between \texttt{BOBE} evidence estimate and the analytic evidence reference value (green dashed line), plotted as a function of the number of BO iterations.  The red dashed line denotes the convergence criterion.  \textbf{Bottom Row}: Grid evaluations of the true function (left), \texttt{BOBE} GP predictive mean (centre) and their difference (right). The run was initialised with 2 Sobol samples.  \label{fig:gaussian}}
\end{figure}

\begin{figure}
    \centering
    \includegraphics[width=0.95\linewidth]{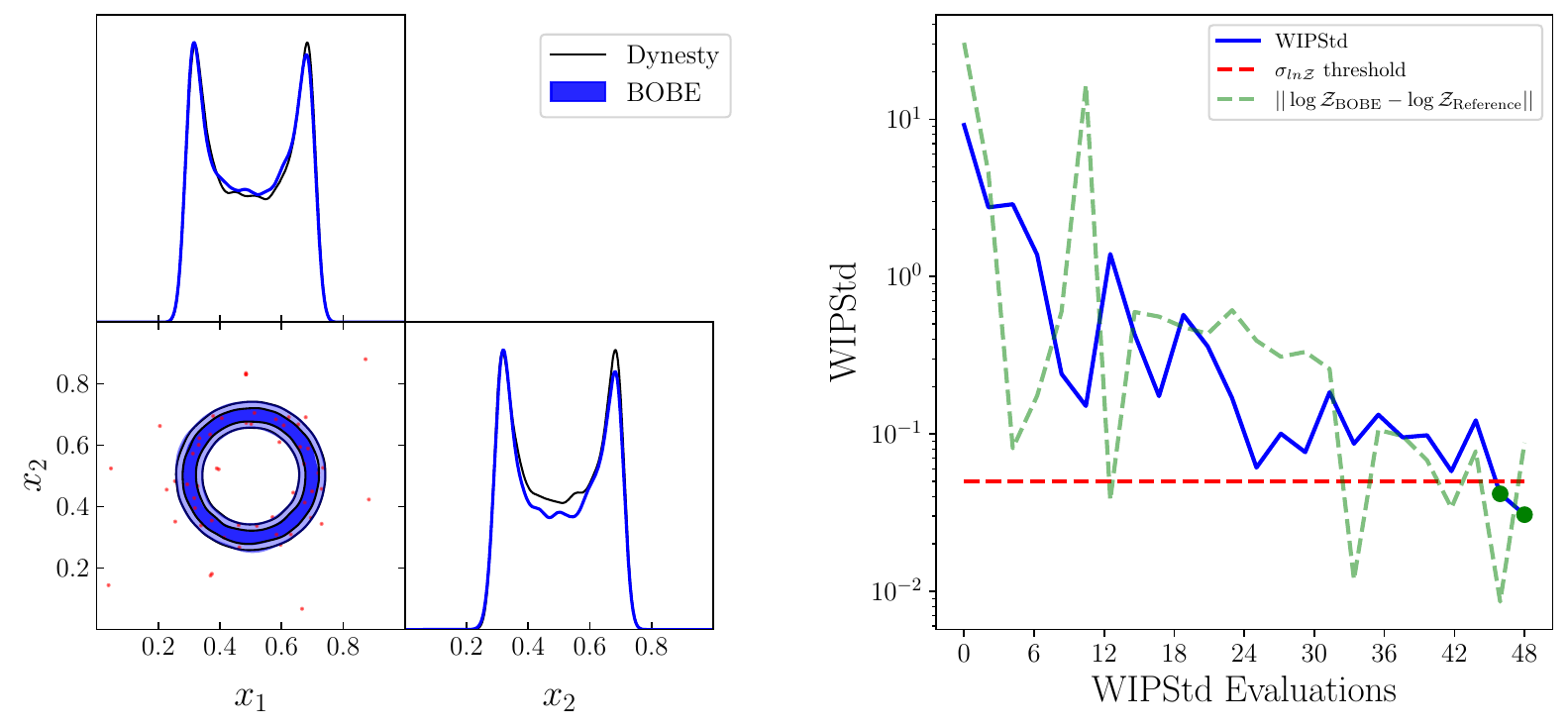}
    \includegraphics[width=0.95\linewidth]{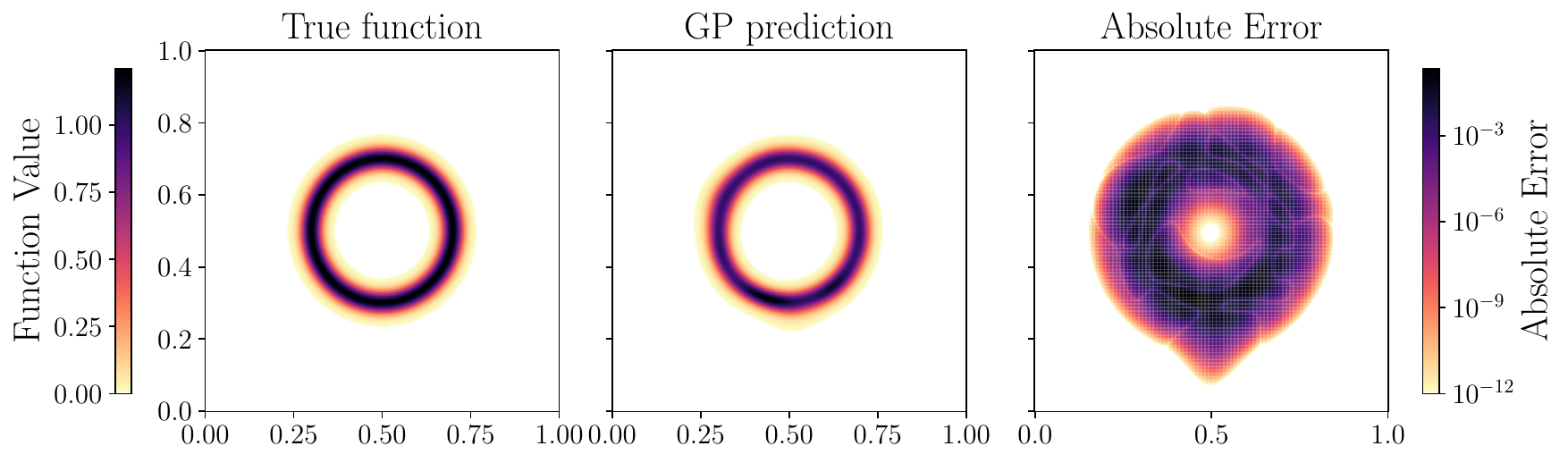}
    \caption{Same as Figure~\ref{fig:gaussian} for the Gaussian ring.  This run was initialised with 8~Sobol samples.
    \label{fig:gaussian_ring}}
\end{figure}

\begin{figure}
    \centering
    \includegraphics[width=0.95\linewidth]{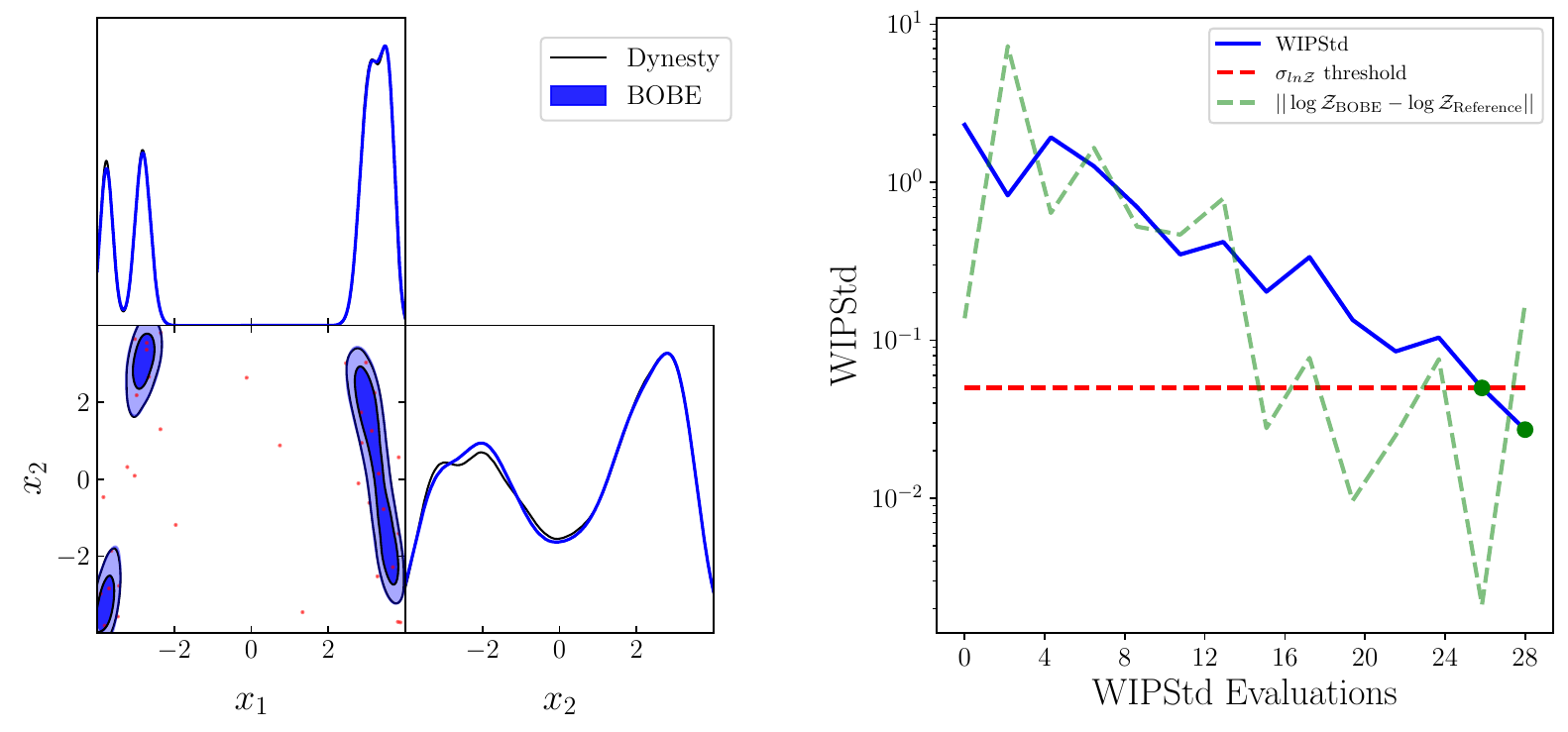}
    \includegraphics[width=0.95\linewidth]{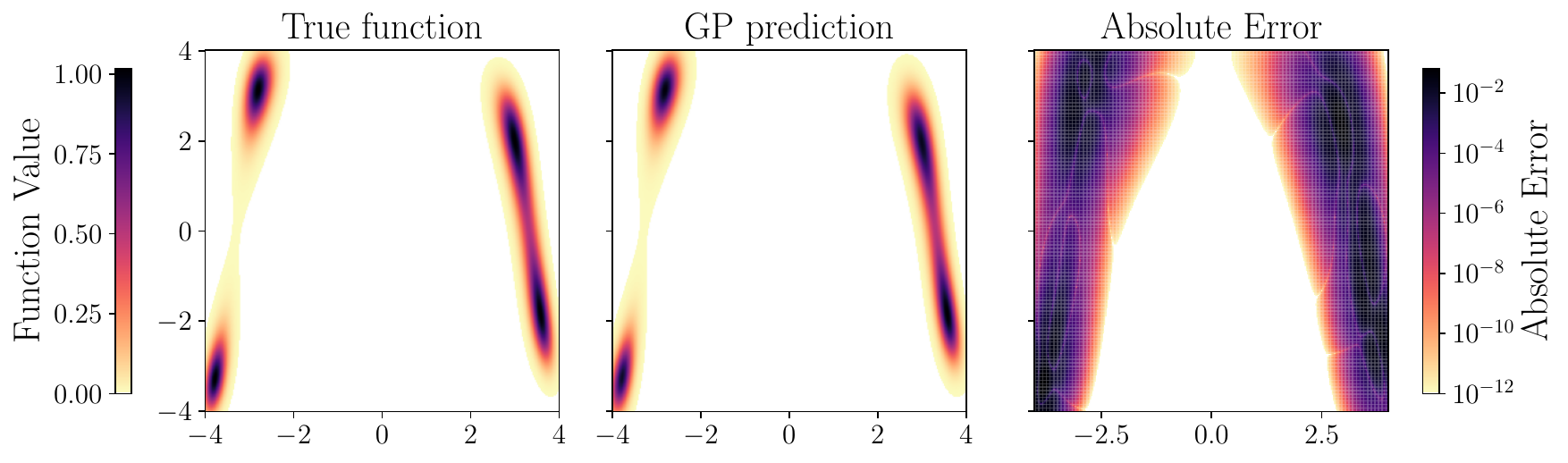}
    \caption{Same as Figure~\ref{fig:gaussian} for the Himmelblau function.  This run was initialised with 8~Sobol samples.
    \label{fig:himmelblau}}
\end{figure}

\begin{figure}
    \centering
    \includegraphics[width=0.95\linewidth]{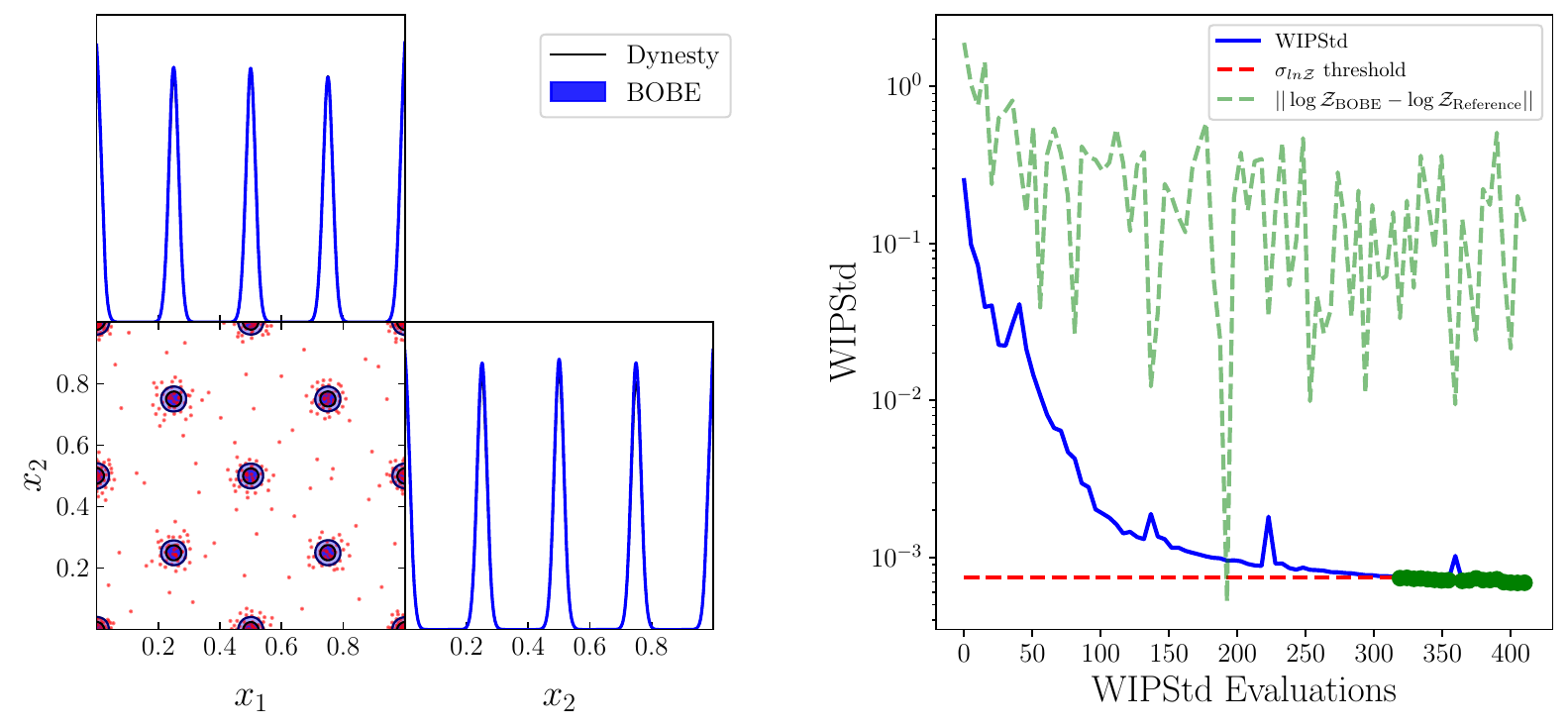}
    \includegraphics[width=0.95\linewidth]{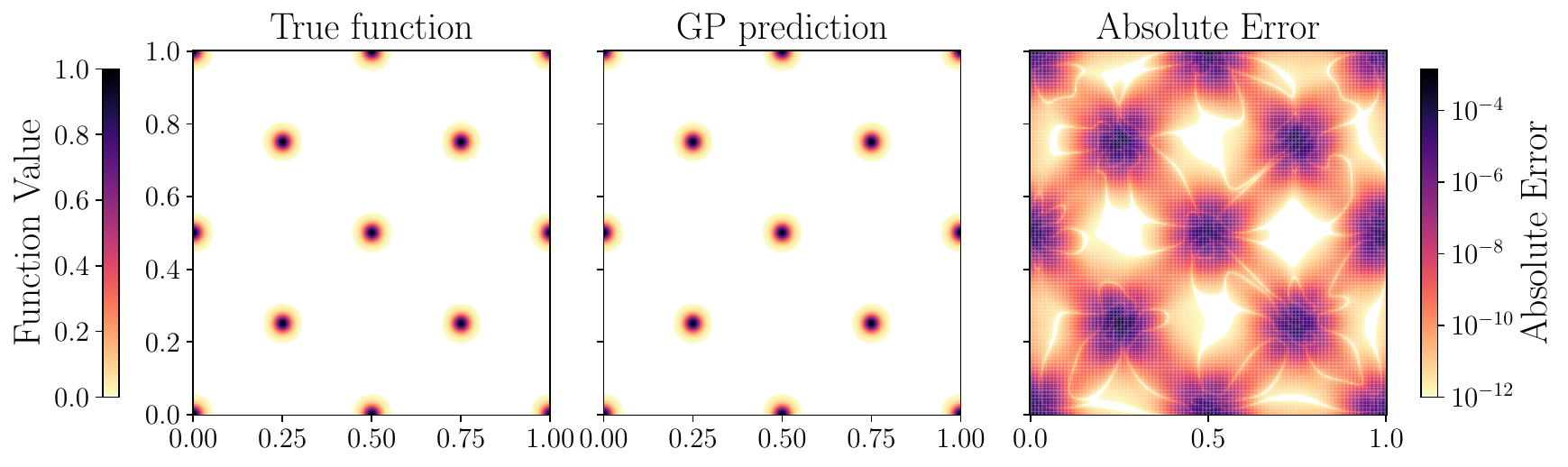}
    \caption{Same as Figure~\ref{fig:gaussian} for the Eggbox function.  This run was initialised with 64~Sobol samples.
    \label{fig:eggbox}}
\end{figure}

We present our results of applying \texttt{BOBE} on these functions in Figures~\ref{fig:gaussian}, \ref{fig:gaussian_ring}, \ref{fig:himmelblau} and \ref{fig:eggbox} \red{using the settings outlined in Table~\ref{tab:BOBE2dsettings}}.  

\begin{table}[h!]
    \centering
    \begin{tabular}{|c|c|c|c|c|c|c|} \hline
     \textbf{Function} & \makecell{Gaussian\\($D \leq 2$)} & \makecell{Gaussian\\($2 < D \leq 4$)} &  \makecell{Gaussian\\($D > 4$)} & \makecell{Gaussian\\ring} & 
    Himmelblau & Eggbox \\ \hline
    Sobol samples & 2 & 4 & 8 & 8 & 8 & 64 \\ 
    Batch size & 1 & 1 & 1 & 2 & 2 & 5 \\ 
    \shortstack{$n_\mathfrak{W}$} & 128 & 128 & 128 & 512 & 512 & 512 \\ 
    \shortstack{$\tau_{\mathrm{refit}}$} & 4 & 4 & 4 & 1 & 1 & 1 \\ 
    Classifier & SVM & SVM & SVM & None & None & None \\ 
    Convergence threshold & $0.005$ & $0.005$ & $0.005$ & $0.05$ & $0.05$ & $0.00075$ \\ 
    $n_\mathrm{iters}$ & 2 & 2 & 2 & 2 & 2 & 10\\ \hline
    \end{tabular}\\
    \caption{\texttt{BOBE} settings for 2D test function runs.}
    \label{tab:BOBE2dsettings}
\end{table}

There are a few points worth mentioning:
\begin{itemize}
    \item{With the exception of the eggbox, WIPStd tracks quite closely the actual difference between the \texttt{BOBE} evidence estimate and the analytically computed ground truth, showing that it is indeed a well-calibrated measure of error.  Note that the difference to the ground truth is affected by both emulator error (which on average decreases as we take more samples) and integrator error (which has a constant expectation value of $\sim 0.1$.)  The total error therefore saturates around 0.1, as clearly visible in Figure~\ref{fig:eggbox}.}
    \item{In all four examples, the final \texttt{BOBE} estimate of the evidence agrees with the true evidence within the integration error.}
    \item{The eggbox proves to be a particularly challenging problem, as all 13 modes need to be discovered for an accurate result.  We therefore started with a particularly large number of Sobol samples (so that some, but not all, of the modes are known initially), and chose much stricter convergence criteria, to force the code to continue running until all modes are detected and explored.}
    \item{In all four examples, it is apparent that the ``interesting'' parts of the functions (that have either high posterior values or a large gradient) are much more densely sampled than the regions of low posterior.}
    \item{\red{Even though these functions are less amenable to modelling with a simple Gaussian kernel than the multivariate Gaussian, they still require substantially fewer likelihood evaluations than direct nested sampling integration, with reductions of approximately $300$, $500$ and $100$ times in the number of samples required for the Gaussian ring, Himmelblau and eggbox functions, respectively.}}
    \item{\rev{The absolute errors of the emulated posteriors are lowest in the low-posterior regions.  This is not surprising, since \texttt{BOBE} emulates $\ln \hat{\mathcal{P}}$, so with the absolute emulation error on the log-posterior given by $\Delta \ln \hat{\mathcal{P}}$, the corresponding absolute error on the posterior is \mbox{$\Delta \hat{\mathcal{P}} \sim \hat{\mathcal{P}} \, \Delta \ln \hat{\mathcal{P}}$}, i.e., it is approximately proportional to the posterior itself.}}
    \item{There is excellent agreement between the reconstructed 1D and 2D posteriors obtained from sampling the \texttt{BOBE} emulator and the true posterior, demonstrating that \texttt{BOBE} can be used for inference as well.}
\end{itemize}

\red{The eggbox example demonstrates that \texttt{BOBE} is able to perform well even for tricky multimodal functions, but we do note that this did require us to tweak its settings.  In principle, if left to run indefinitely (and if not using a classifier), \texttt{BOBE} would eventually discover all isolated maxima by virtue of the design of WIPStd and its innate ability to switch back and forth from exploitation to exploration.  However, if the convergence criterion is satisfied before all relevant maxima of the posterior are mapped out, \texttt{BOBE} will underestimate both the evidence and the error. 

In cases where a user suspects a target function to have multimodal behaviour, we recommend several practical methods of safeguarding premature convergence when the user suspects the target function to be multimodal:
\begin{itemize}
    \item Increase the chance of having discovered the relevant modes before commencing WIPStd sampling – for instance by increasing the number of initial Sobol samples or by running \texttt{BOBE} with a more exploration-focused acquisition function before switching to WIPStd.
    \item Delay convergence, e.g., by tightening the WIPStd convergence threshold or increasing $n_{\mathrm{iters}}$.
    \item Forego use of the classifier, as once a region has been excluded, discovery of modes contained within becomes impossible.
\end{itemize}
Lastly, as with any non-deterministic sampling method, one could also run \texttt{BOBE} multiple times with different random seeds.}

\subsection{\textit{\textbf{N}}-dimensional Gaussian: scaling with dimensionality}
Next we investigate how \texttt{BOBE}'s performance scales with the dimensionality of parameter space.  Motivated by the fact that multivariate near-Gaussian posteriors are common in cosmological contexts, we consider for this analysis an $N$-dimensional Gaussian posterior,
\begin{align}
    \ln \hat{\mathcal{P}}_{\mathrm{G}-N\mathrm{D}}(\boldsymbol{x}) = -\frac{1}{{2\sigma^2}} \,{\displaystyle \sum_{i=1}^{N} (x_i - \mu)^2}
\end{align}
with $\mu=0.5$, $\sigma=0.1$, $x_i \in \{ 0, 1\}$, and run \texttt{BOBE} until a target precision of $\sigma_{\ln \mathcal{Z}} \approx \widehat{\mathfrak{W}} = 0.1$ is reached. \red{The remaining detailed settings are shown in Table~\ref{tab:BOBE2dsettings}, but we note that performance depends only very weakly on these settings here}.  

The focus here will in particular be on the total number of function calls to the posterior required for \texttt{BOBE} to converge (as minimising this quantity was one of our design goals).  We compare this with how many function calls would be required if one instead were to follow the standard way of computing evidences, i.e., by using a nested sampler (\texttt{Dynesty}, with an integration precision of $\sigma_{\ln \mathcal{Z}} = 0.1$) directly calling the likelihood.

\begin{figure}
    \centering
    \includegraphics[width=0.95\linewidth]{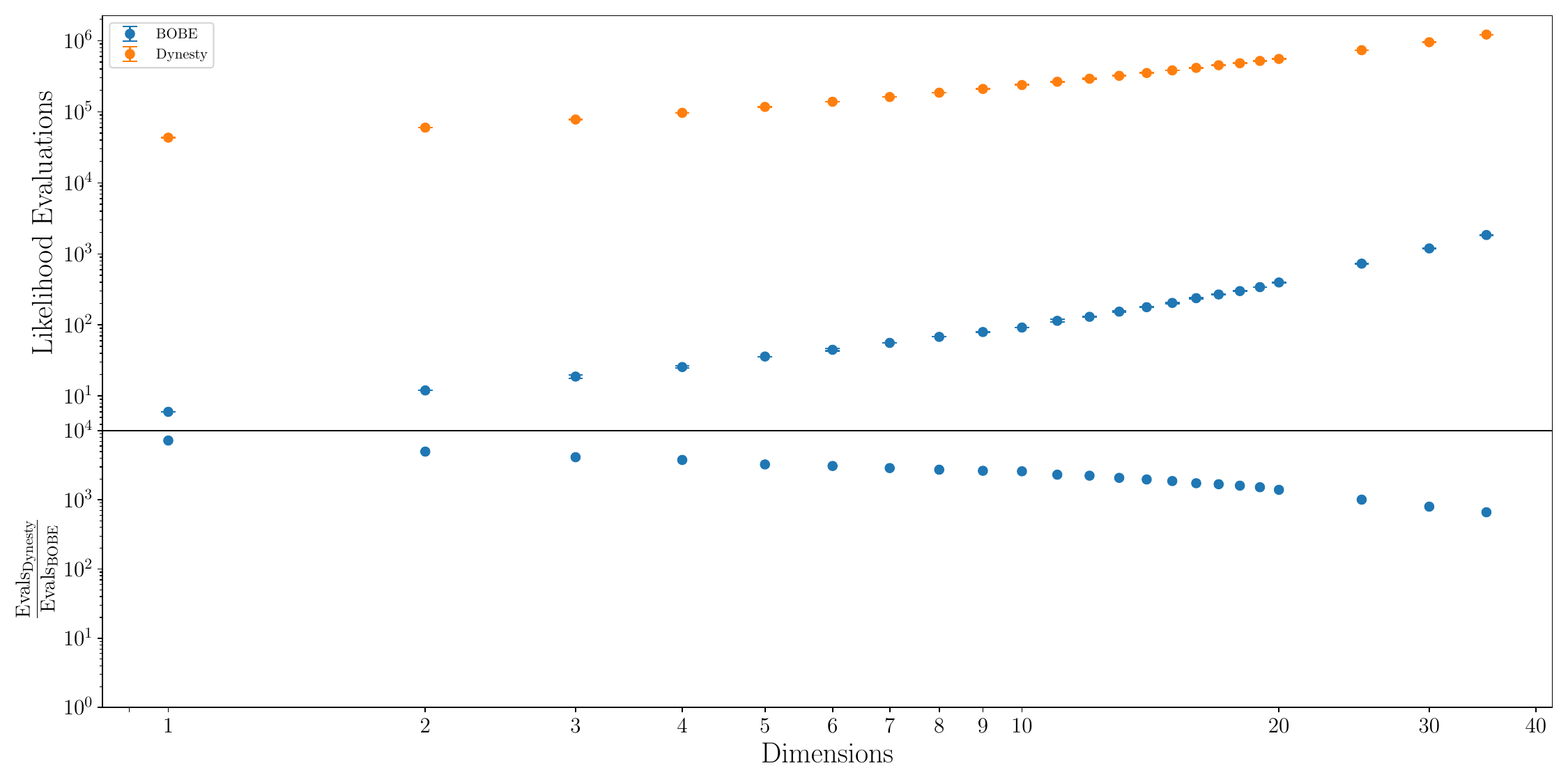}
    \caption{\textbf{Top:} Number of calls to an $N$-dimensional Gaussian likelihood/posterior required for \texttt{BOBE} to be trained (blue), compared with \texttt{Dynesty}, for an emulation/integration precision $\sigma_{\ln \mathcal{Z}} = 0.1$.  \textbf{Bottom:} Ratio of the two numbers.
    \label{fig:BOBESamplesvsDimension}}
\end{figure}

Our results for the number of posterior calls up to $N = 35$ dimensions are presented in Figure~\ref{fig:BOBESamplesvsDimension}.  It is clear that building the \texttt{BOBE} emulator is substantially more efficient by this metric than nested sampling with direct calls to the posterior, requiring $\mathcal{O}(10^3)$ fewer evaluations.  
Since \texttt{BOBE} does have a larger computational overhead than nested samplers, this does not directly translate to the same factor in terms of computational time required (though in the limit of sufficiently expensive likelihoods it will\footnote{If one is concerned with overall computation time, the breakeven point in terms of a single likelihood evaluation time $t_\mathcal{L}$ depends on the computing hardware and $N$ – in our examples it lies around $t_\mathcal{L} \sim 10^{-2}$ to $10^{-1}$~s, i.e., for likelihoods slower than this, \texttt{BOBE} will be faster.}).

Both \texttt{BOBE} and \texttt{Dynesty} require more function calls with increasing $N$. \rev{We do note that in this particular example, \texttt{BOBE}'s scaling is slightly less favourable, but a rough extrapolation of the trend in Figure~\ref{fig:BOBESamplesvsDimension} suggests that it may continue to enjoy a higher efficiency well beyond $N \sim 100$ dimensions for multidimensional Gaussian posteriors.}

\begin{figure}
    \centering
    \includegraphics[width=0.75\linewidth]{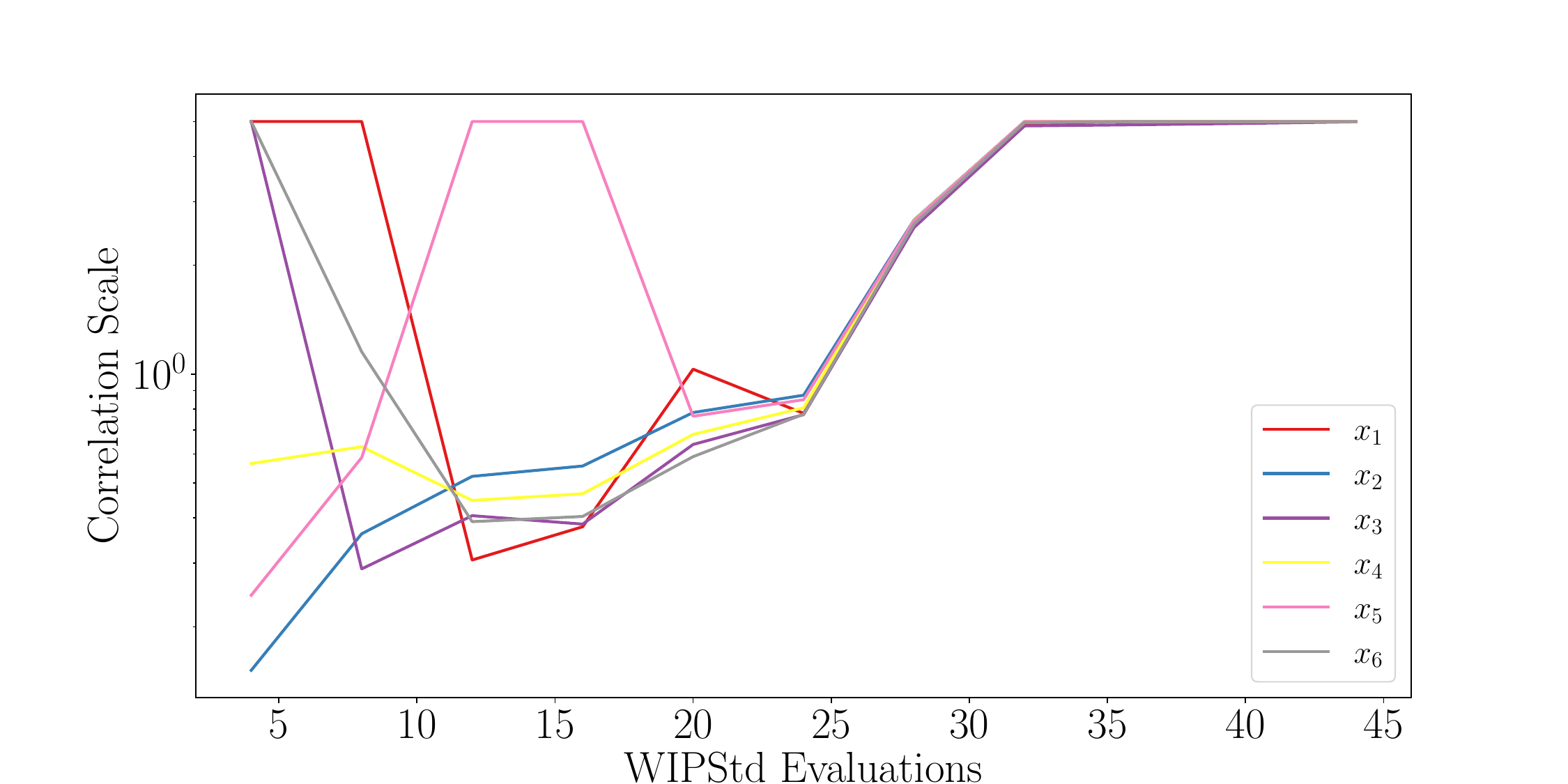}
    \caption{Evolution  of correlation scale hyperparameters as a function of iteration from a \texttt{BOBE} run on a 6D Gaussian.}
    \label{fig:hp_plot}
\end{figure}

In the context of the $N$-dimensional Gaussian case, it may also be enlightening to inspect the behaviour of the GP hyperparameters during the training process.  We illustrate this for the example of a 6D Gaussian in Fig \ref{fig:hp_plot}.  The evolution of the correlation scales reveals two distinct regimes.  An initial phase, characterised by strong fluctuations and marked differences between length scales for the different parameters, during which the rough shape of the posterior is being explored.  This is followed by a second phase, where the different length scales have converged to a common value (as demanded by the symmetry of the problem), indicating a transition to an exploitation regime, i.e., the refinement of the emulator mostly in high-posterior regions of parameter space.

\subsection{A cosmological application}
\label{subsec:cosmo_examples}
Finally, we apply \texttt{BOBE} on an actual cosmological example, namely the base 6-parameter $\Lambda$CDM model, as well as its extension with non-zero spatial curvature, $\Lambda$CDM+$\Omega_k$.  We consider the following likelihoods through their interface with \texttt{Cobaya}~\cite{Torrado:2020dgo}:
\begin{itemize}
    \item{\textbf{CMB-lite}: a combination of the \textit{Planck} 2018 low-$\ell$ temperature and polarisation likelihood~\cite{Planck:2019nip}, the \textit{Planck-lite} nuisance-marginalised high-$\ell$ TTTEEE temperature and polarisation likelihood~\cite{Planck:2019nip} and the \textit{Planck} 2018 lensing likelihood~\cite{Aghanim:2018oex}.} 
    \item{\textbf{CMB}: as above, but replacing \textit{Planck-lite} with \textit{Planck} high-$\ell$ \texttt{CamSpec} TTTEEE temperature and polarisation likelihood using \texttt{NPIPE} (\textit{Planck} PR4) data~\cite{Rosenberg:2022sdy}, which adds 9 nuisance parameters.} 
    \item{\textbf{BAO}: Baryon Acoustic Oscillation (BAO) data from data release 2 of the DESI collaboration~\cite{DESI:2025zgx}.} 
\end{itemize}
The prior ranges imposed on cosmological and nuisance parameters are listed in Table~\ref{tab:priors_planck_omk} in Appendix~\ref{sec:full_cosmo_posteriors}. Cosmological observables required as input by these likelihoods are computed using the Boltzmann code \texttt{CAMB}~\cite{Lewis:1999bs}.  

We use a 2023 Apple M2 Pro with 10 cores and 32 GB of RAM for our computation. Thanks to parallelisation, the 10 cores can be efficiently used in their entirety, and \texttt{BOBE}'s memory requirements typically do not exceed 2-3~GB, even at a GP training set size of 1000\texttt{+}.  All \texttt{BOBE} runs use batch acquisition and are initialised with a training set containing a combination of Sobol samples and samples drawn from a \texttt{Cobaya} reference distribution, with the details given in Table~\ref{tab:BOBEcosmosettings}.  

\begin{figure}
    \centering
    \includegraphics[width=0.5\linewidth]{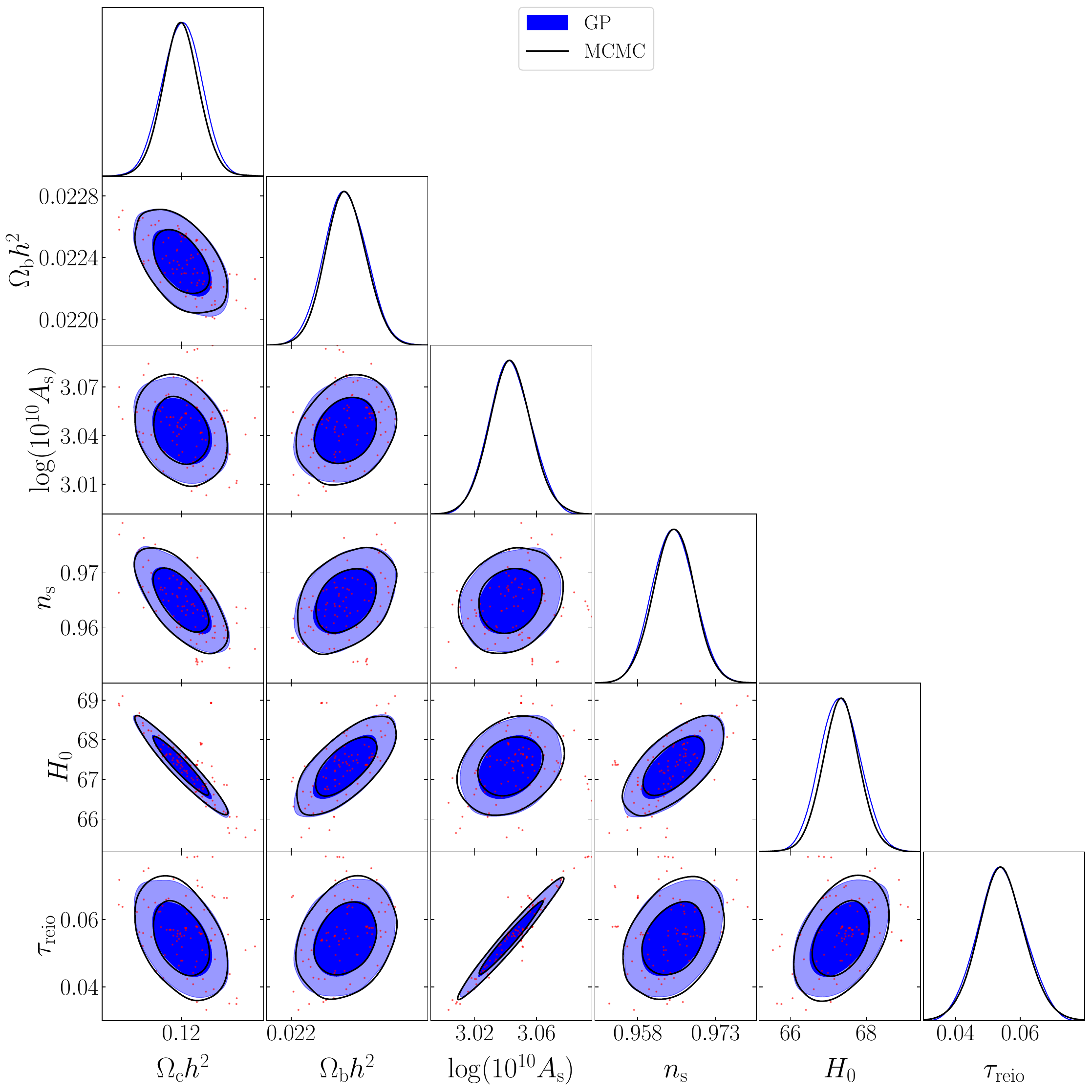}
    \includegraphics[width=0.49\linewidth]{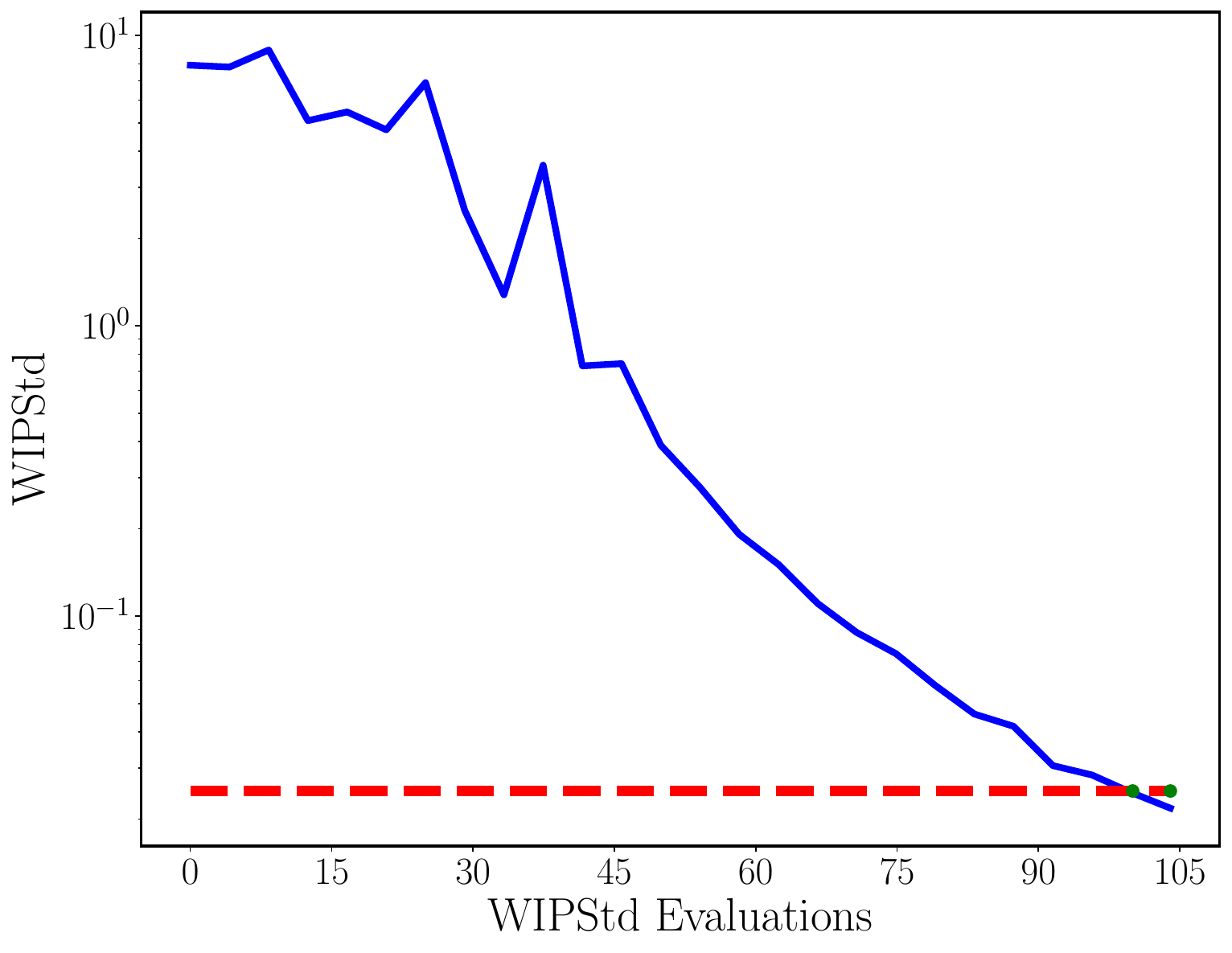}
    \caption{\textit{Left}: Cosmological parameter posteriors for the $\Lambda$CDM model using the CMB-lite likelihood from the trained GP emulator (blue blobs/lines) compared to results from running MCMC on the actual likelihood (black lines).  We also show the locations of training samples as red dots. \textit{Right}: Evolution of WIPStd convergence diagnostic (blue line). The red dashed line denotes the convergence criterion. 
    \label{fig:lcdm_lite}}
% \end{figure}

% \begin{figure}
%     \centering
\bigskip
\centering
    \includegraphics[width=0.49\linewidth]{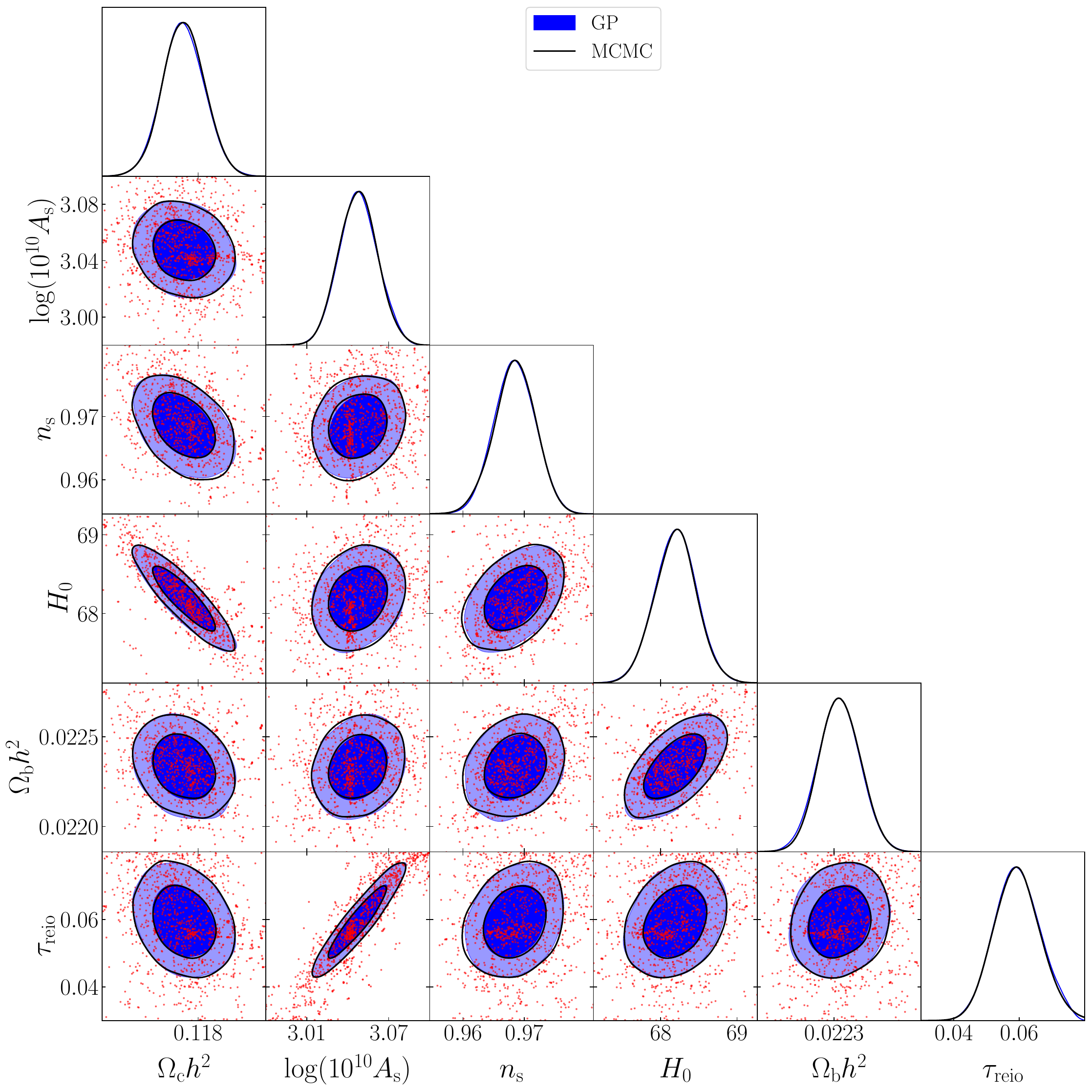}
    \includegraphics[width=0.49\linewidth]{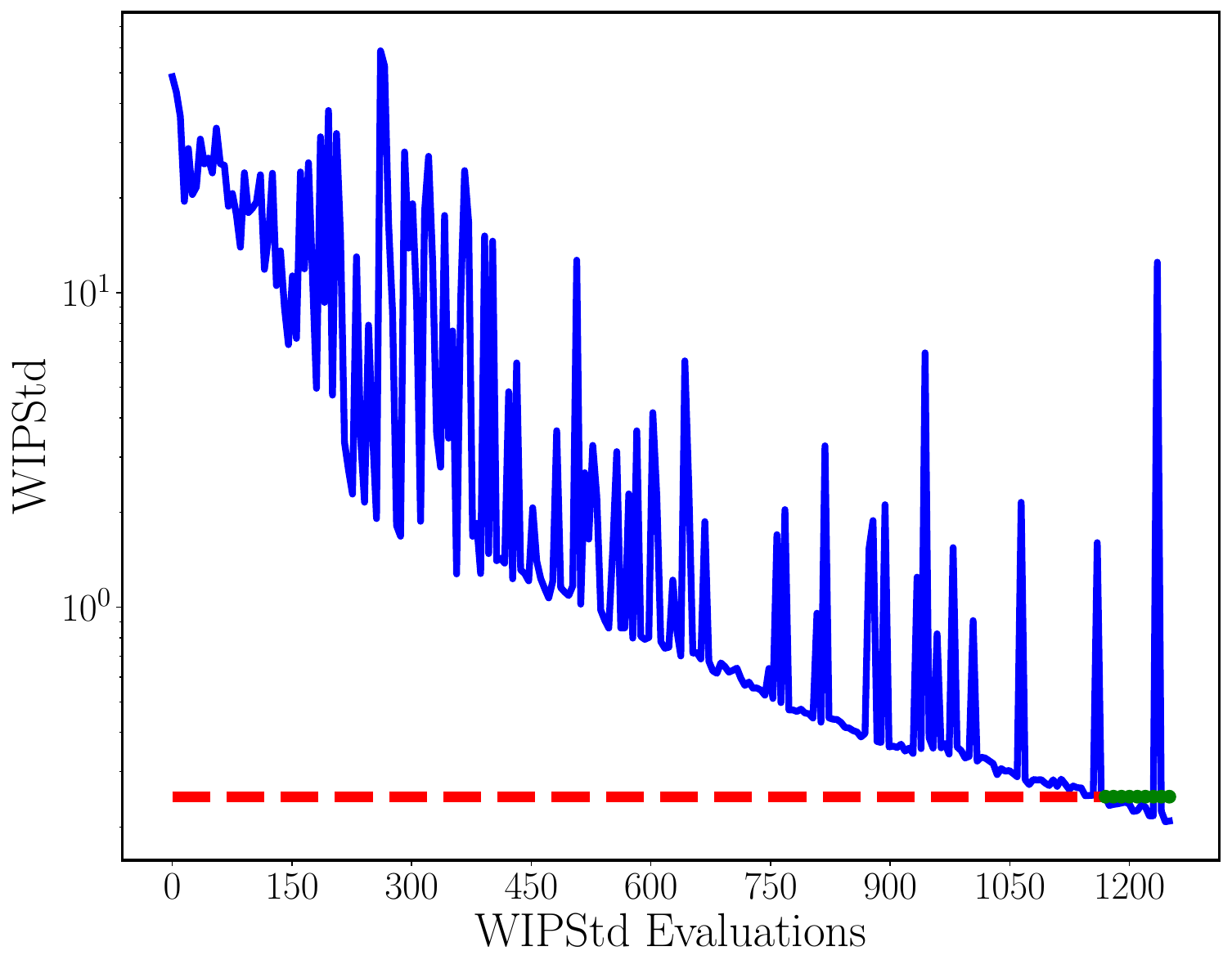}
    \caption{Same as Figure~\ref{fig:lcdm_lite} for the $\Lambda$CDM model using the CMB+BAO likelihoods. 
    \label{fig:lcdm_planck_desi}}
\end{figure}

\begin{figure}
    \centering
    \includegraphics[width=0.49\linewidth]{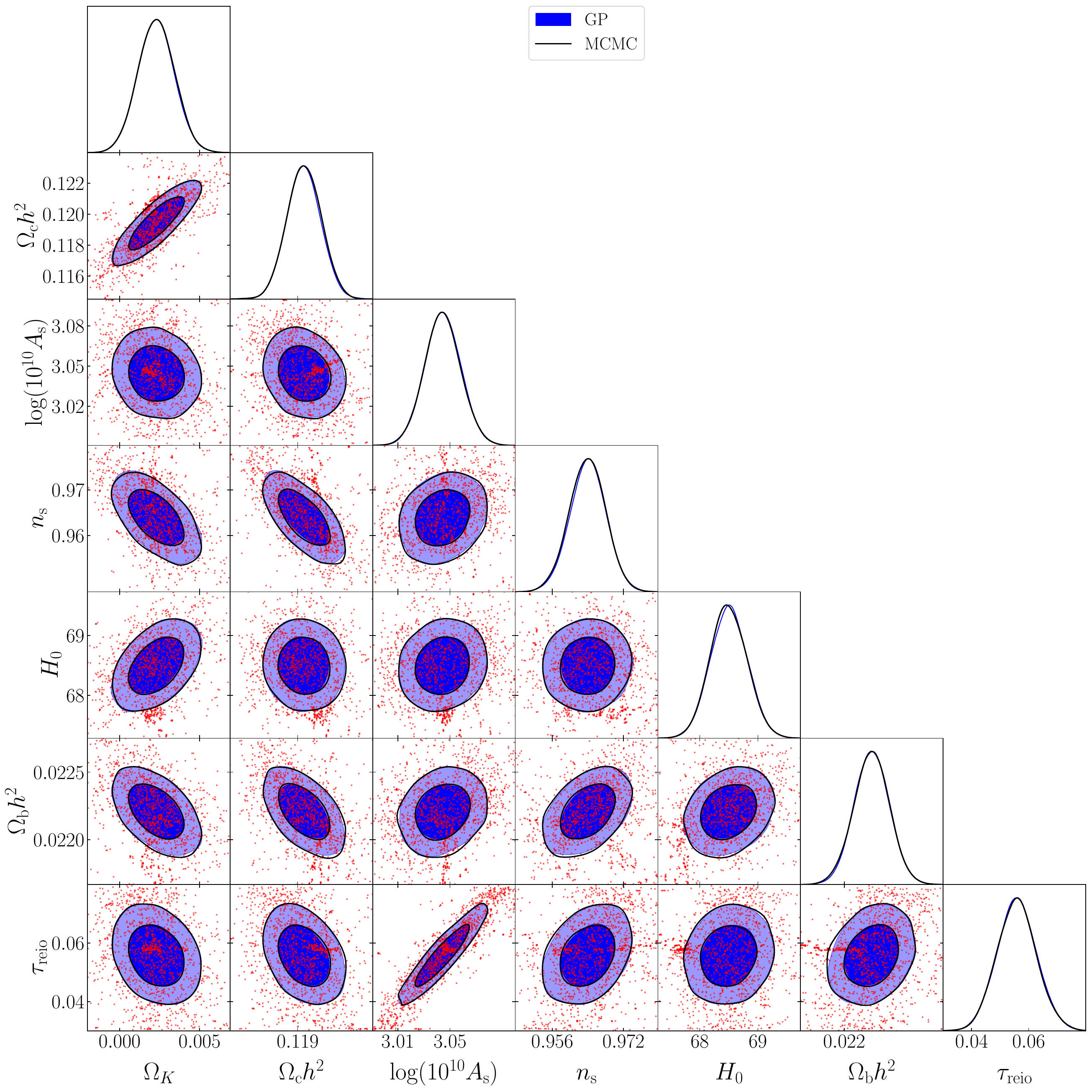}
    \includegraphics[width=0.49\linewidth]{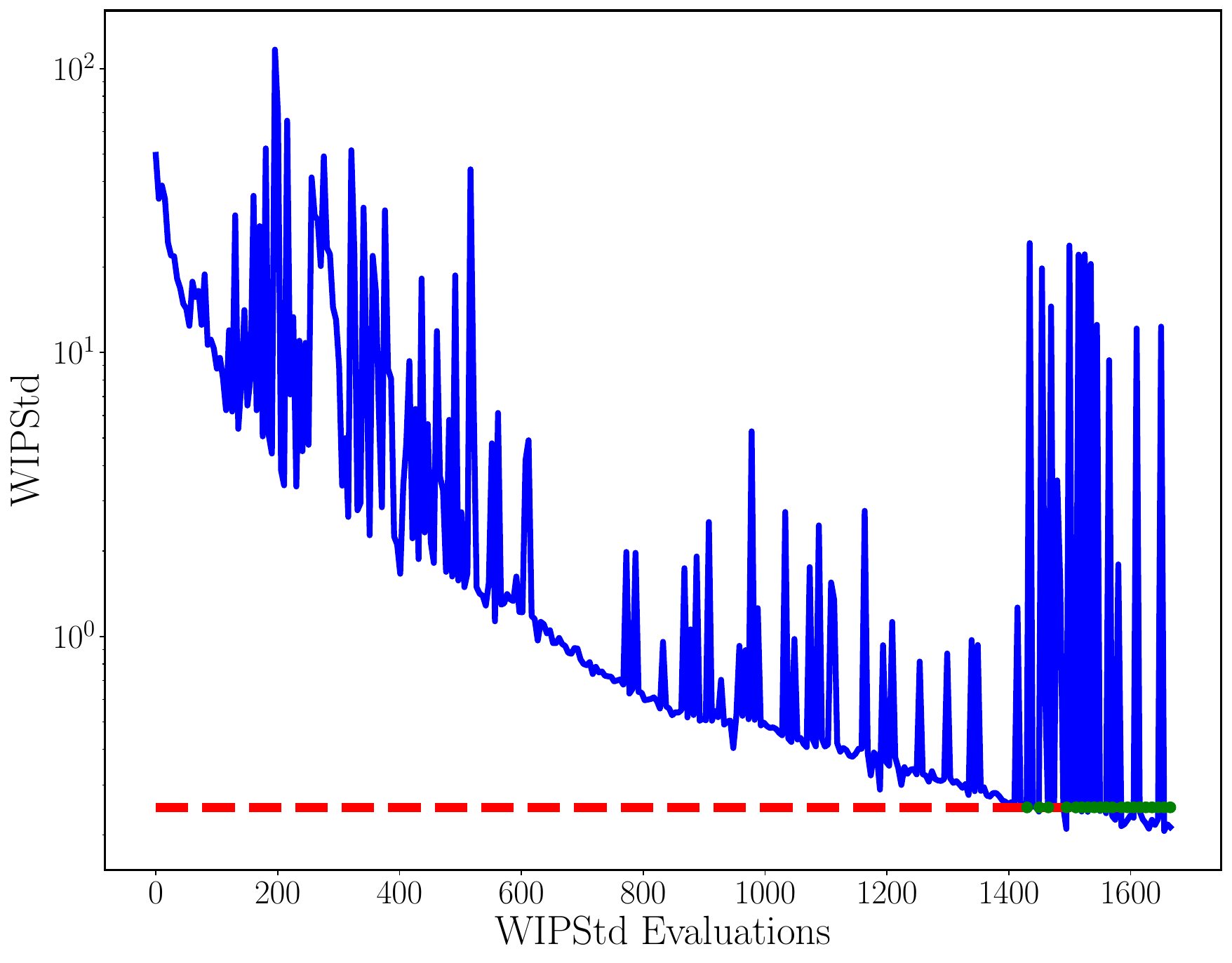}
    \caption{Same as Figure~\ref{fig:lcdm_lite} for the $\Lambda$CDM+$\Omega_k$ model using the CMB+BAO likelihoods.
    \label{fig:lcdm_curvature}}
\end{figure}

The convergence history and inferred 1- and 2-dimensional marginalised posteriors for the cosmological parameters are shown in Figure~\ref{fig:lcdm_lite} for $\Lambda$CDM~+~CMB-lite, in Figure~\ref{fig:lcdm_planck_desi} for $\Lambda$CDM~+~CMB+BAO and in Figure~\ref{fig:lcdm_curvature} for $\Lambda$CDM+$\Omega_k$~+~CMB+BAO (see also Figures~\ref{fig:LCDM_Planck_DESI_all} and~\ref{fig:LCDM_Omk_Planck_DESI_all} in Appendix~\ref{sec:full_cosmo_posteriors} for full posteriors including nuisance parameters). In all cases, the posteriors obtained from the \texttt{BOBE} emulator match those obtained from a nested sampling run on the actual likelihoods, demonstrating \texttt{BOBE}'s suitability for inference problems.

For comparison, we also compute evidences with \texttt{PolyChord} (which is much faster than other nested samplers on the full CMB likelihood, due to its ability to exploit the difference between fast and slow parameters), using the actual likelihoods.

We list the \texttt{PolyChord} evidence estimates along with the results of our \texttt{BOBE} evidence computation in Table~\ref{tab:BOBEcosmoresults} – there is excellent agreement between the two for all combinations of model/data, but \texttt{BOBE} only took a fraction of the time to run on the same hardware.  As already suggested by Figure~\ref{fig:BOBESamplesvsDimension}, \texttt{BOBE}'s advantage is particularly pronounced in the lower-dimensional $\Lambda$CDM~+~CMB-lite case (3~min vs. 10~hours), but still substantial in the higher-dimensional problems involving the full CMB likelihood (54~min vs. $\sim 1$~day, and 2.5~hours vs.~more than a day). \red{We provide the settings we use for these runs in Table \ref{tab:BOBEcosmosettings}.}

\begin{table}[h!]
    \centering
    \begin{tabular}{|c|c|c|c|} \hline
    \textbf{Model} & $\Lambda$CDM & $\Lambda$CDM & $\Lambda$CDM+$\Omega_k$ \\ \hline
    Datasets & CMB-lite & CMB+BAO & CMB+BAO \\ 
    Sobol samples & 8 & 32 & 32 \\ 
    \texttt{Cobaya} samples & 4 & 8 & 16 \\ 
    Batch size & 4 & 5 & 5 \\ 
    \shortstack{$n_\mathfrak{W}$} & 512 & 512 & 512 \\ 
    \shortstack{$\tau_{\mathrm{refit}}$} & 4 & 5 & 5 \\ 
    Classifier & SVM & SVM & SVM \\ 
    Convergence threshold & $0.025$ & $0.5$ & $0.5$ \\ 
    $n_\mathrm{iters}$ & 2 & 2 & 2 \\ \hline
    \end{tabular}\\
    \caption{\texttt{BOBE} settings for cosmology runs. 
    \label{tab:BOBEcosmosettings}}
\bigskip
\centering
    \begin{tabular}{|c|c|c|c|c|c|} \hline
    Model & Data & Samples & Wall time & $\ln \zc \pm \sigma_\mathrm{ns} \pm \sigma_\texttt{BOBE}$~ (\texttt{BOBE}) & $\ln \zc$ (\texttt{PolyChord})\\ \hline
    $\Lambda$CDM & CMB-lite & 116 & 3 min & $-519.84 \pm 0.15 \pm 0.02$ & $-520.18 \pm 0.32$\\
    $\Lambda$CDM & CMB+BAO & 1213 & 54 min & $-5529.84 \pm 0.2 \pm 0.2$ & $-5529.65 \pm 0.45$\\
    $\Lambda$CDM+$\Omega_k$ & CMB+BAO & 1713 & 2.5 hours & $-5529.93 \pm 0.24 \pm 0.22$ & $-5529.80 \pm 0.37$\\
    \hline
    \end{tabular}
    \caption{Results of \texttt{BOBE} cosmology runs. \rev{We quote both the nested sampler uncertainty ($\sigma_\texttt{ns}$) and the emulation uncertainty ($\sigma_\texttt{BOBE}$, given simply by the value of $\widehat{\mathfrak{W}}$ at convergence) for the \texttt{BOBE} evidence estimate.}
    \label{tab:BOBEcosmoresults}}
\end{table}

\clearpage

\section{Conclusion}
\label{sec:conclusion}

In this work, we have presented a Bayesian Optimisation framework for efficient Bayesian evidence estimation in cosmology and beyond, addressing the computational challenge of expensive likelihood evaluations. By combining Gaussian Process surrogates with our Weighted Integrated log-Posterior Standard Deviation (WIPStd) acquisition function, our method achieves accurate evidence estimates with up to three orders of magnitude fewer likelihood evaluations compared to traditional approaches. Further, the byproduct of efficient parameter inference makes this algorithm an even more attractive option for expensive likelihoods. 

By making use of an evidence aware strategy that ensures likelihood evaluations are guided by their impact on the precision of the evidence estimate, we have significantly reduced the number of true likelihood evaluations required to approximate the evidence to a given precision. The uncertainty quantification capabilities of GPs provide a natural convergence criteria based on the reduction of posterior-weighted standard deviation while the added combination of dynamic range compression through the choice of log-likelihood emulation and classifier guided exploration yields a robust and scalable framework. Future work could extend this framework by using richer Gaussian process constructions such as additive \cite{additive_GP}, sparse \cite{sparse_GP} random Fourier feature~\cite{additive_rff_GP}, deep kernels \cite{deep_GP} or integrate multi-fidelity modelling \cite{MF_GP} to accommodate even higher dimensional parameter spaces.

We have applied our algorithm to various problems across different dimensionalities, from simple Gaussian likelihoods to multimodal, non-Gaussian test examples and most importantly realistic cosmological inference problems including $\Lambda$CDM and its extension with spatial curvature. Our framework can efficiently handle low to moderate-dimensional ($\lesssim 20$D) parameter spaces, achieving significant efficiency gains over traditional Bayesian methods based on MCMC or nested sampling. This makes Bayesian model selection practical for computationally expensive scenarios where traditional sampling methods become prohibitive or even intractable. Overall, the combination of principled uncertainty quantification, active learning, and evidence-aware acquisition using BO provides a powerful tool for rigorous model comparison.

\subsection*{Acknowledgements}
AM is partially funded by the STFC grant ST/X000648/1. AM also acknowledges the support of the Supercomputing Wales project, which is part-funded by the European Regional Development Fund (ERDF) via Welsh Government. NC acknowledges support from an Australian Government Research Training Program Scholarship. This research includes computations using the computational cluster Katana supported by Research Technology Services at UNSW Sydney. For the purpose of open access, the authors have applied a Creative Commons Attribution (CC BY) licence to any Author Accepted Manuscript version arising. Research Data Access Statement: our code, along with the scripts used to run the examples presented in this manuscript can be obtained from \url{https://github.com/Ameek94/BOBE}.

\begin{appendix}
\section{Classifier implementation \label{sec:CLF}}
The classifier ensures that the emulator remains both accurate and tractable in high-dimensional settings. To set the log-likelihood threshold where we draw the classifier boundary, we follow the strategy used in \texttt{Gpry} \cite{Gammal:2022eob}. This threshold strategy considers a posterior that is locally Gaussian in the region around the posterior maximum $\boldsymbol{x}_0$ in a $d$-dimensional parameter space. Under this assumption the posterior in the region $\boldsymbol{x}_0$ can be written as
\begin{align}
    p(\boldsymbol{x}) \propto \exp\left[-\frac{1}{2}(\boldsymbol{x} - \boldsymbol{x}_0)^{\intercal}\mathbf{C}^{-1}(\boldsymbol{x} - \boldsymbol{x}_0) \right],
\end{align}
where $C$ is the covariance matrix. The quadratic form can then be defined as
\begin{align}
    \Delta(\boldsymbol{x}) \equiv (\boldsymbol{x} - \boldsymbol{x}_0)^{\intercal}\mathbf{C}^{-1}(\boldsymbol{x} - \boldsymbol{x}_0),
\end{align}
which follows a $\chi^2$ distribution with $N_d$ degrees of freedom,
\begin{align}
    \Delta \sim \chi^2_{N_d}.
\end{align}
In one dimension a deviation of $n_{\sigma}$ corresponds to a change in probability mass
\begin{align}
    p = \mathrm{erfc}\left(\frac{n_{\sigma}}{\sqrt{2}} \right).
\end{align}
In higher dimensions, the corresponding confidence region must enclose the same probability mass, which is achieved by requiring
\begin{align}
    P(\Delta \leq \Delta_d) = 1 - p\quad \mathrm{and} \quad \Delta_d = \chi^2_{d, \mathrm{isf}}(p)
\end{align}
where $\chi^2_{d, \mathrm{isf}}$ is the $\chi^2$ inverse survival function with $d$ degrees of freedom. So, an effective $n_{\sigma}$ confidence region in $d$ dimensions is defined by
\begin{align}
    \Delta_{d}(n_{\sigma}) =  \chi^2_{d, \mathrm{isf}} \left(\mathrm{erfc}\left(\frac{n_{\sigma}}{\sqrt{2}} \right) \right).
    \label{eq:grpy_thresh}
\end{align}
This models the fact that the scale of an equivalent confidence region grows with dimensionality and, in the case of our algorithm, enables an approximation of what threshold from the peak is relevant to estimation of the evidence. 

In our implementation we consider three complementary approaches, all of which draw a boundary relative to the current GP posterior maximum to determine admissibility using Equation~\eqref{eq:grpy_thresh}. The first uses a Support Vector Machine (SVM) through the \texttt{scikit-learn} library~\cite{scikit-learn} as also done in \texttt{Gpry}~\cite{Gammal:2022eob}. In addition to the SVM classifier, we implement a neural network based classifier using \texttt{Flax}~\cite{flax2020github}, as well as a classifier that draws an ellipsoidal boundary centred on the current GP posterior maximum point. When the posterior is expected to be Gaussian or uni-modal, the ellipsoidal classifier provides the most accurate representation of its shape. For more complex posteriors, the SVM or neural network based classifiers are preferable as their greater flexibility allows them to adapt to more intricate posterior structures. In practice, the SVM offers the best performance on our hardware and is therefore selected as the default choice.

\section{Parameter priors and posterior plots}
\label{sec:full_cosmo_posteriors}

\begin{table}[h!]
    \centering
    \begin{tabular}{lc}
    \hline
    Parameter & Prior  \\
    \hline
    $\Omega_k$ & [-0.010, 0.010] \\
    $\Omega_\mathrm{c} h^2$ & [0.11, 0.13] \\
    $\Omega_\mathrm{b} h^2$ & [0.021, 0.023] \\
    $\ln(10^{10}A_{\mathrm{s}})$ & [2.98, 3.10] \\
    $n_{\mathrm{s}}$ & [0.94, 0.99] \\
    $H_0 \: [\mathrm{km}\: \mathrm{s^{-1}}\: \mathrm{Mpc^{-1}}]$ & [62, 72] \\
    $\tau_{\mathrm{reio}}$ & [0.03, 0.08] \\ 
    \noalign{\vskip 2pt}
    \hdashline 
    \noalign{\vskip 2pt}
    $A^{\rm power}_{143}$ & [0, 50] \\
    $A^{\rm power}_{217}$ & [0, 50] \\
    $A^{\rm power}_{143 \times 217}$ & [0, 50] \\
    $\gamma^{\rm power}_{143}$ & [0, 5] \\
    $\gamma^{\rm power}_{217}$ & [0, 5] \\
    $\gamma^{\rm power}_{143 \times 217}$ & [0, 5] \\
    $A_{\mathrm{planck}}$ & [0.9875, 1.0125] \\
    $c_{TE}$ & [0.950, 1.050] \\
    $c_{EE}$ & [0.950, 1.050] \\
    \hline
    \end{tabular}
    \caption{Prior ranges for the cosmological parameters of the $\Lambda$CDM and $\Lambda$CDM+$\Omega_k$ models, as well as for the nuisance parameters required by the \textit{Planck} \texttt{CamSpec} likelihood (below the dashed line). Prior ranges are chosen to be wide enough to remain non-informative relative to the posterior, while being narrower than typical MCMC parameter inference priors in order to ensure that the nested sampling algorithm converges within a reasonable number of steps.
    \label{tab:priors_planck_omk}}
\end{table}

\begin{figure}
    \centering
    \includegraphics[width=0.95\linewidth]{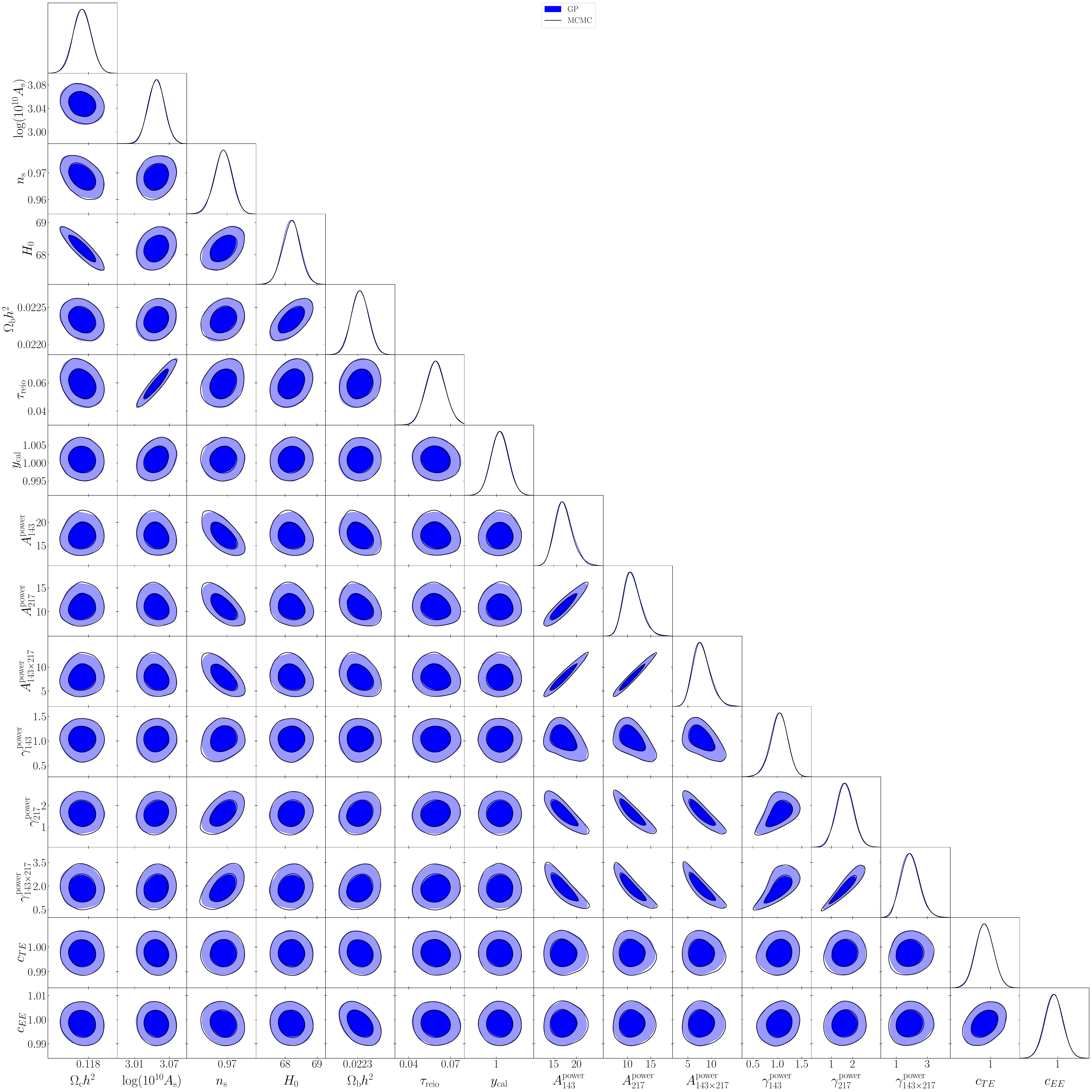}
    \caption{Cosmological and nuisance parameter posteriors for the $\Lambda$CDM model using the CMB+BAO likelihood from the trained GP emulator (blue blobs/lines) compared to results from running MCMC on the actual likelihood (black lines). 
    \label{fig:LCDM_Planck_DESI_all}}
\end{figure}

\begin{figure}
    \centering
    \includegraphics[width=0.95\linewidth]{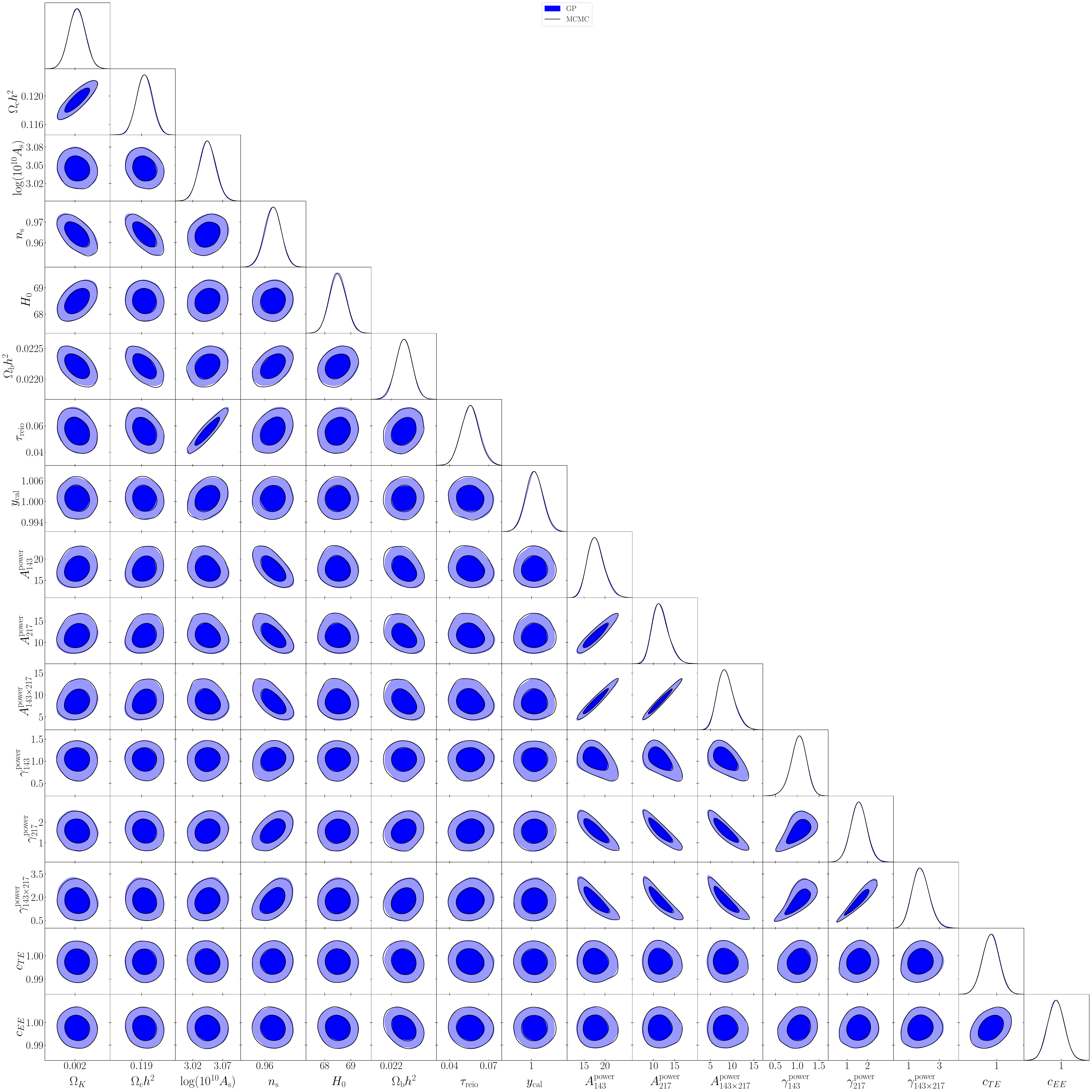}
    \caption{Same as Figure~\ref{fig:LCDM_Planck_DESI_all} for the $\Lambda$CDM+$\Omega_k$ model.
    \label{fig:LCDM_Omk_Planck_DESI_all}}
\end{figure}

\end{appendix}
\clearpage
\bibliographystyle{JHEP}
\bibliography{bibliography}

\end{document}